\documentclass[sigconf]{static/package/acmart}
\usepackage{array,enumitem,diagbox,amsmath,amsthm,booktabs,comment,graphicx,subfigure,multirow,bm,tikz,makecell,xcolor,ulem,hyperref,marvosym,filecontents,xparse,tcolorbox}
\usepackage[ruled,linesnumbered]{algorithm2e}
\usepackage{pdfpages}

\usetikzlibrary{shapes,decorations}
\definecolor{bule}{RGB}{18,29,57}
\definecolor{bablue}{RGB}{248,248,248}
\definecolor{main}{RGB}{127,191,51}
\definecolor{seco}{RGB}{0,145,215}
\definecolor{thid}{RGB}{180,27,131}
\definecolor{Gainsboro}{RGB}{200,200,200}
\definecolor{DimGrey}{RGB}{105,105,105}
\newcommand{\newfancytheoremstyle}[5]{
  \tikzset{#1/.style={draw=#3, fill=#2, very thick, rectangle,
      rounded corners, inner sep=10pt, inner ysep=10pt}}
  \tikzset{#1title/.style={fill=#3, text=#2}}
  \expandafter\def\csname #1headstyle\endcsname{#4}
  \expandafter\def\csname #1bodystyle\endcsname{#5}
}
\makeatletter
\NewDocumentCommand{\newfancytheorem}{O{\@empty} m m m O{fancythrm} }{
  \ifx#1\@empty
    \newcounter{#2}
  \else
    \newcounter{#2}[#1]
    \numberwithin{#2}{#1}
  \fi
  \NewEnviron{#2}[1][{}]{
    \par \vspace{5pt} \noindent\centering
    \begin{tikzpicture}
      \node[#5](box){
        \begin{minipage}{0.9\columnwidth}
          \csname #5bodystyle\endcsname \BODY
        \end{minipage}
      };
      \node[#5title, right=10pt] at (box.north west){
        {\csname #5headstyle \endcsname #3 \stepcounter{#2}\csname the#2\endcsname\; ##1}
      };
    \end{tikzpicture}
  }[\par\vspace{.5\baselineskip}]
}
\makeatother
\newfancytheoremstyle{fancythrm}{Gainsboro!10}{seco}{\bfseries\sffamily}{\sffamily}
\newfancytheoremstyle{theoremsty}{Gainsboro!10}{DimGrey}{\itshape\sffamily}{\sffamily}
\newfancytheorem{mydef}{Definition}{$\spadesuit$}[theoremsty]
\newcommand{\defname}[1]{\textbf{\textcolor{white}{#1}}}
\newcommand{\todo}[1]{\textbf{\textcolor{red}{#1}}}

\setcopyright{none}
\acmYear{2025}
\definecolor{cbrickred}{rgb}{0.8, 0.25, 0.33}
  
\newcommand{\brickred}[1]{{\color{cbrickred} #1}}

\newcommand{\update}[1]{}
\newcommand{\partitle}[1]{\smallskip \noindent \textbf{#1.}}

\newcommand{\projectname}{{\textsc{ControlNet}}}
\newcolumntype{M}[1]{>{\centering\arraybackslash}m{#1}}  
\newcolumntype{R}[1]{>{\raggedright\arraybackslash}m{#1}} 

\begin{document}
\title{\projectname: A Firewall for RAG-based LLM System}
\author{\normalsize{Hongwei Yao$^{1,2}$, Haoran Shi$^{2}$, Yidou Chen$^{2}$, Yixin Jiang$^{3}$, Cong Wang$^{1}$\Letter, Zhan Qin$^{2}$\Letter}}
\affiliation{
  \normalsize{$^{1}$City University of Hong Kong, Hong Kong China} \\
  \normalsize{$^{2}$Zhejiang University, Hangzhou China} \\
  \normalsize{$^{3}$China Southern Power Grid, Guangzhou China} \\
  Project page: \href{https://ai.zjuicsr.cn/firewall}{https://ai.zjuicsr.cn/firewall} \country{}\\
}

\begin{abstract}
Retrieval-Augmented Generation (RAG) has significantly enhanced the factual accuracy and domain adaptability of Large Language Models (LLMs). This advancement has enabled their widespread deployment across sensitive domains such as healthcare, finance, and enterprise applications. RAG mitigates hallucinations by integrating external knowledge, yet it introduces privacy risk and security risk, notably \textbf{data breaching risk} and \textbf{data poisoning risk}. While recent studies have explored prompt injection and poisoning attacks, there remains a significant gap in comprehensive research on controlling inbound/outbound query flows to mitigate these threats. 
In this paper, we propose an AI firewall, \projectname, designed to safeguard RAG-based LLM systems from these vulnerabilities. \projectname~ controls query flows by leveraging activation shift phenomena to detect malicious queries and mitigate their impact through semantic divergence. We conduct comprehensive experiments using state-of-the-art open-source LLMs—Llama3, Vicuna, and Mistral—across four benchmark datasets (MS MARCO, HotpotQA, FinQA, and MedicalSys). Our empirical results demonstrate that \projectname~ is not only effective but also harmless. It achieves an \texttt{AUROC} exceeding 0.909 for risk detection, with minimal degradation in \texttt{Precision} and \texttt{Recall}, both of which show reductions of less than 0.03 and 0.09, respectively, for risk mitigation. Overall, \projectname~ offers an effective, harmless, robust defense mechanism, marking a significant advancement toward the secure deployment of RAG-based LLM systems.
\end{abstract}

\maketitle

\section{Introduction}
In recent years, Large Language Models (LLMs) such as ChatGPT, Claude and DeepSeek~\cite{guo2025deepseek} have demonstrated remarkable proficiency in domain-specific and knowledge-intensive tasks, including personal assistant, medical assistant, financial advisor, and legal research. This advancement has been significantly driven by the integration of Retrieval-Augmented Generation (RAG) techniques. 
By incorporating external knowledge retrieval, RAG mitigates the hallucination problem inherent in standalone generative models~\cite{YeLZHJ24,GunjalYB24}, improving their ability to respond to queries requiring real-time or out-of-distribution information~\cite{fan2024survey}.

\begin{figure}[!t]
    \centering
    \includegraphics[width=0.9\linewidth]{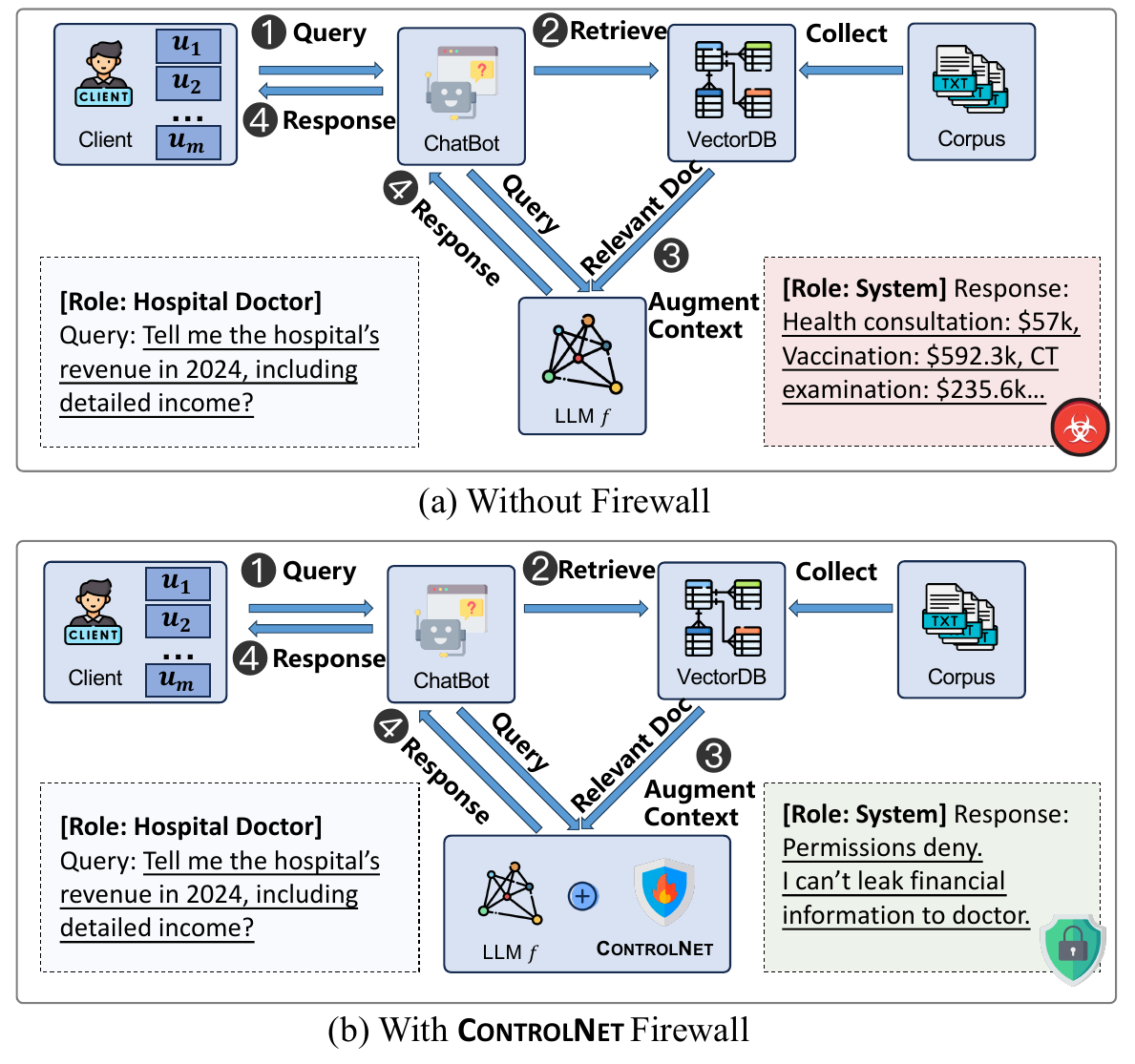}
    \caption{Illustration of the data flow in a RAG-based LLM system. 
    (a) Without the firewall, the doctor gains unauthorized access to financial data. (b) With the firewall \projectname, role-based access control ensures the doctor can only retrieve patient information.}
    \label{fig:rag_data_flow}
\end{figure}

A typical RAG-based LLM system operates by processing user queries through a structured retrieval mechanism. When a client submits a query, the server encodes it into a vector representation, which is then used to retrieve relevant documents from a vector database (denoted as \texttt{VectorDB}). These documents, stored in vector form, are subsequently integrated with the original query and processed by the LLM to generate a response. The knowledge provider ensures that the \texttt{VectorDB} remains updated with accurate and relevant documents. 
Finally, the generated response is delivered to the client (as shown in Figure~\ref{fig:rag_data_flow} (a)). 
This retrieval-based approach has proven highly effective in reducing factual inaccuracies, contributing to the widespread adoption of RAG as a foundational technology for enhancing the practical applicability of LLMs.

Despite these advancements, the incorporation of external knowledge documents introduces critical vulnerabilities related to \textbf{privacy} and \textbf{security}. 
These risks manifest in two primary forms, \textbf{data breaching risk}~\footnote{\href{https://securityintelligence.com/articles/chatgpt-confirms-data-breach}{https://securityintelligence.com/articles/chatgpt-confirms-data-breach}} and \textbf{data poisoning risk}~\footnote{\href{https://www.theguardian.com/technology/2024/dec/24/chatgpt-search-tool-vulnerable-to-manipulation-and-deception-tests-show}{https://www.theguardian.com/technology/2024/dec/24/chatgpt-search-tool-vulnerable-to-manipulation-and-deception-tests-show}}.
First, data breaching risks arise from malicious clients conducting reconnaissance to extract system prompts or exploit RAG system environments. In multi-client systems where users have differentiated access rights (e.g., executives, financial officers, general staff), such vulnerabilities increase the likelihood of data exfiltration and unauthorized document access(as shown in Figure~\ref{fig:rag_data_flow} (b)). 
Data breaching can lead to the exposure of system environments and business-sensitive information, posing severe consequences for organizations.
Second, from the corpus collection perspective, data poisoning presents another security challenge. Malicious knowledge providers may introduce compromised data into the \texttt{VectorDB}, thereby injecting misleading information or hijacking clients’ conversation.
Data poisoning can mislead clients, degrade trust in the system, and compromise decision-making processes.
Given these escalating risks, there is an urgent need for effective, robust defense mechanisms to ensure the secure and trustworthy deployment of RAG-based LLM systems.

Ensuring the integrity and security of RAG-based LLM systems requires the development of effective mechanisms for detecting malicious queries, filtering poisoned documents, and implementing effective query flow control.
While recent studies have explored prompt injection~\cite{liu2024formalizing,liu2023prompt,shi2024optimization,zhu2024promptbench} and poisoning attacks~\cite{yao2024poisonprompt,yang2024tapi,zhan2024injecagent,yu2023assessing,ToyerWMSBWOEADR24}, and guardrails~\cite{inan2023llama,LlamaGuard2,RebedeaDSPC23} in LLMs, 
there remains a significant gap in comprehensive research on controlling inbound/outbound query flows to mitigate these threats.
Although guardrails contribute to safety and alignment but often fail to manage complex, multi-role system flows effectively.
Addressing these risks in RAG-based systems presents several key challenges.
First, both incoming queries and retrieved documents exist as unstructured textual data, rendering traditional regular expression-based matching ineffective. Instead, effective flow control should leverage deeper semantic features, such as neuron activation patterns within the model. Second, establishing a principled connection between neuron activation patterns and client-based access control in RAG systems remains an open problem. Third, once an malicious query or poisoned document is detected, mitigating their impact through query sanitization presents a further challenge. 
To address these challenges, future research must prioritize the design of robust query sanitization techniques and adversarial impact mitigation strategies. 

\partitle{Our Work} 
In this paper, we introduce a comprehensive RAG security framework that considers three distinct entities: the LLM server, the clients, and the corpus collector. We systematically investigate privacy and security risks in RAG-based LLM systems, identifying five major attacks: \textit{reconnaissance, data exfiltration, unauthorized access, knowledge poisoning, and conversation hijacking}. To address these threats, we propose \projectname, a novel AI firewall for RAG-based LLM systems. \projectname~ controls query flow by leveraging activation shift phenomena, detecting malicious queries based on distinct activation vector patterns indicative of semantic divergence. Upon detection, \projectname~ mitigates risks by steering the LLM’s behavior away from harmful responses, thereby ensuring secure and privacy-preserving interactions.
We conduct extensive experiments using three state-of-the-art (SoTA) open-source LLMs—Llama3, Vicuna, and Mistral—across four benchmark datasets: MS MARCO, HotpotQA, FinQA, and MedicalSys.
These datasets cover diverse application scenarios, including personal assistant, digital enterprise, financial planning, and healthcare.
Our empirical results substantiate that \projectname~ is not only effective but also harmless and robust. It achieves an \texttt{AUROC} exceeding 0.909 for risk detection, with minimal degradation in \texttt{Precision} and \texttt{Recall}, both of which show reductions of less than 0.03 and 0.09, respectively, for risk mitigation.

The contributions of this work are summarized as follows:
\begin{itemize}[leftmargin=14pt]
    \item \textbf{Systematic Risk Analysis}: 
    We present a comprehensive taxonomy of privacy and security vulnerabilities in RAG-based systems, categorizing five attacks under two primary risk domains. This framework provides a foundational structure for future research and development.
    \item \textbf{Benchmark Dataset}:
    We collect and release an open-access, multi-role database for the medical system with local hospitals, comprising over 20,000 data samples across 4 roles. This database serves as a benchmark for evaluating privacy risks and fostering further research.
    \item \textbf{Next-Generation AI Firewall}:
    We propose \projectname, a novel AI firewall framework for query flow control in RAG-based LLM systems. \projectname~ controls query flow by leveraging activation shift phenomena, detecting malicious queries through distinct activation vector patterns indicative of semantic divergence.
    \item \textbf{Extensive Experiments}:
    Through extensive experiments on SoTA LLMs and benchmark datasets, we demonstrate \projectname’s effectiveness in detecting and mitigating privacy and security risks, ensuring reliable deployment of RAG systems in real-world applications.
\end{itemize}

\section{Related Works}
\subsection{LLM Attacks}
In recent years, the security risks associated with LLMs have garnered significant attention within the research community.
These risks can be broadly categorized into two main areas: \textbf{privacy threats} and \textbf{security threats}~\cite{esmradi2023comprehensive,yao2024survey,DongZYSQ24}. 
This section delves into existing literature on these topics, particularly focusing on data leakage and prompt injection.

\subsubsection{Data Leakage}
The massive datasets used to train LLMs often inadvertently include sensitive information, such as Personally Identifiable Information (PII), e.g., names, age, email addresses, physical addresses.
This inclusion raises significant privacy concerns, as LLMs are capable of memorizing sensitive information during training and reproducing it during inference.
Early research demonstrated the feasibility of extracting sensitive text sequences from LLMs via membership inference attacks~\cite{DBLP:conf/uss/CarliniTWJHLRBS21,DBLP:conf/uss/CarliniHNJSTBIW23,liu2024mitigating}.
Subsequent studies further investigated the risks associated with PII leakage in LLMs~\cite{DBLP:conf/sp/LukasSSTWB23,zhou2024quantifying,kim2024propile,liu2024evaluating}.

Beyond the leakage of training data, the confidentiality of system prompts has emerged as a critical concern, particularly with the increasing adoption of LLM-driven applications~\cite{WangYXD24,yao2024promptcare}. 
Several high-profile incidents, such as the leakage of the Bing Chat system prompt~\footnote{\url{https://github.com/elder-plinius/Bing-Prompt-Leak}} and similar breaches involving OpenAI's GPTs~\footnote{\url{https://github.com/friuns2/Leaked-GPTs}}, have underscored the vulnerabilities in protecting these assets.
In response, academic research addresses the issue of prompt leakage through the identification of malicious prompts and the development of strategies to mitigate their impact~\cite{DBLP:journals/corr/abs-2402-12959,DBLP:conf/ccs/0002Y0BC24,DBLP:conf/uss/ShenQ0024}.

\subsubsection{Prompt Injection}
Prompt injection represents a significant challenge in the realm of LLMs, involving the unauthorized manipulation of input to alter model behavior, potentially leading to unintended outputs~\cite{liu2024formalizing,liu2023prompt,shi2024optimization}. 
The prompt injection attacks can be categorized into two main types: \textbf{direct injection} and \textbf{indirect injection} attacks~\cite{zhu2024promptbench}.
(1) Direct injection attacks are characterized by adversarial inputs specifically designed to manipulate LLM output, often bypassing intended functionality or ethical boundaries. Various strategies have been reported in literature, including naive attacks, context ignoring~\cite{perez2022ignore}, fake completions~\cite{willison2024delimiters,zou2024poisonedrag}, and more nuanced methods like Do Anything Now (DAN)~\cite{shen2024anything} and the Grandma Exploit~\cite{heuser2024grandma}. 
Each of these approaches exploits different facets of LLMs, from their syntactic processing capabilities to their contextual understanding mechanisms.
(2) Indirect injection attacks, on the other hand, involve manipulating external resources that LLMs rely upon for information.
By altering documents, websites, or database entries, adversaries can influence the data consumed by LLMs, thereby affecting their outputs without directly interacting with the model's prompt interface~\cite{yu2023assessing}. 
Recent studies have highlighted the vulnerability of LLM-integrated applications to such attacks, e.g., \textsc{InjectAgent}~\cite{zhan2024injecagent}, TensorTrust~\cite{ToyerWMSBWOEADR24}, Indirect Prompt Injection~\cite{AbdelnabiGMEHF23}, and TAPI~\cite{yang2024tapi}.

\subsection{LLM Guardrails}
Guardrails for LLMs are structured, rule-based frameworks designed to function as intermediaries, ensuring that interactions between clients and the models comply with predefined safety guidelines and operational protocols.
Current LLM Guardrails can be categorized into categorized into \textbf{alignment-based guardrails} and \textbf{rule-based guardrails} strategies.

\subsubsection{Alignment-based Guardrails} 
Alignment-based guardrails typically involve fine-tuning LLMs on specific instructions or using reinforcement learning from human feedback (RLHF)\cite{ouyang2022training,bai2022training}
to make LLM align with human values and ethical standards\cite{yaacov2024boosting,jan2024multitask}.
These methods often require substantial computational resources and manual effort which constrains their scalability and hinders their feasibility for real-time and widespread deployment.

\subsubsection{Rule-based Guardrails} 
Rule-based guardrails are designed to moderate and mitigate harmful content at various stages, including the input, output, and activation layers of LLMs\cite{kumar2024watch}. 
Recent advancements have harnessed the contextual understanding capabilities of machine learning to develop classifiers that excel within their predefined categories. Nevertheless, these classifiers often face challenges in generalizing to novel or emerging risks, as exemplified by systems such as Google Ads Content Moderation \cite{qiao2024scaling}, \cite{markov2023holistic}, Perspective API\cite{lees2022new} and LlamaGuard\cite{inan2023llama,LlamaGuard2}.
SoTA guardrails have further evolved to incorporate contextual text control, thereby enhancing the mitigation of harmful content. For instance, NeMo Guardrails \cite{RebedeaDSPC23} and RigorLLM\cite{YuanX0Y0S024} integrate external \texttt{VectorDB}s to restrict the output space of LLMs, ensuring alignment with verified facts and contextual appropriateness. 
Additionally, Adaptive Guardrail~\cite{hu2024adaptive} and PrimeGuard\cite{manczak2024primeguard} introduce a two-step process: initially identifying potentially harmful requests in the prompt, followed by mitigating any unaligned content in the response. 

\subsubsection{Limitations} 
While LLM guardrails contribute significantly to safety and alignment,
they often fall short in controlling multi-role system flows. These limitations highlight the necessity for adaptive and robust AI firewall designs to manage complex user interactions and permissions.

\section{Preliminaries}
\subsection{Large Language Model}
A LLM $f$ is a parametric function that estimates the conditional probability of a sequence of words $w_{1:t} = (w_1, w_2, ..., w_t)$ given a context of previous words, where $w_i \in \mathcal{V}$ and $\mathcal{V}$ is the vocabulary of the model. 
The model is trained on a large corpus $\mathcal{C}$ and is represented by a deep neural network with parameters $\theta$. 
The conditional probability is given by:\update{2025.03.10}
\begin{equation}
P(w_{1:t} | w_{1:t-1}; \theta) = \prod_{i=1}^{t} P(w_i | w_{1:i-1}; \theta),
\end{equation}
where $P(w_i | w_{1:i-1}; \theta)$ is the probability of the $i$-th word given the previous words, computed by the model $f$ with parameters $\theta$.

\subsection{Retrieval-Augmented Generation}
RAG is a technique that combines a retriever component with a generator component. 
The retriever is used to fetch relevant documents from a large knowledge \texttt{VectorDB}, which is then integrated into the generator to produce coherent and informative text. 
Formally, RAG can be defined as a function $\mathcal{R}(q, p; \phi, \theta)$ that takes a query $q$ and a system prompt $p$, and outputs a sequence of words $w_{1:t}$ based on the context.

\partitle{Retriever Component}
Given an incoming query $q$ and a system prompt $p$, the retriever $r$ identifies a set of $k$ relevant documents $\mathcal{D} = \{d_1, d_2, ..., d_k\}$ from \texttt{VectorDB}:
\begin{equation}
    \mathcal{D} = r(q, p; \phi),
\end{equation}
where $\phi$ represents the parameters of the retriever.

\partitle{Generator Component}
The language model $f$ takes the system prompt $p$, the query $q$, and the retrieved documents $\mathcal{D}$ to generate a sequence of words $w_{1:t}$.
\begin{equation}
    w_{1:t} = f(q, p, \mathcal{D}; \theta),
\end{equation}
where $\theta$ represents the parameters of the language model.
The overall RAG model can be expressed as:
\begin{equation}
\mathcal{R}(q, p; \phi, \theta) = f(q, p, r(q, p; \phi); \theta),
\end{equation}
The training objective is to find the parameters $\phi^*$ and $\theta^*$ that maximize the conditional probability of the target sequence $w_{1:t}$ given the system prompt $p$ and query $q$, with the constraint that the retrieved documents are relevant:
\begin{equation}
\phi^*, \theta^* = \arg\max_{\phi, \theta} \sum_{(q, p, w_{1:t}) \in \mathcal{T}} \log P(w_{1:t} | q, p, r(q, p; \phi); \theta),
\end{equation}
where $\mathcal{T}$ is the main task containing triplets of query, context and target response sequences.
\update{2025.03.10}

\section{Risks in RAG-based LLM system}
\begin{figure*}[ht!]
  \centering
  \includegraphics[width=0.9\linewidth]{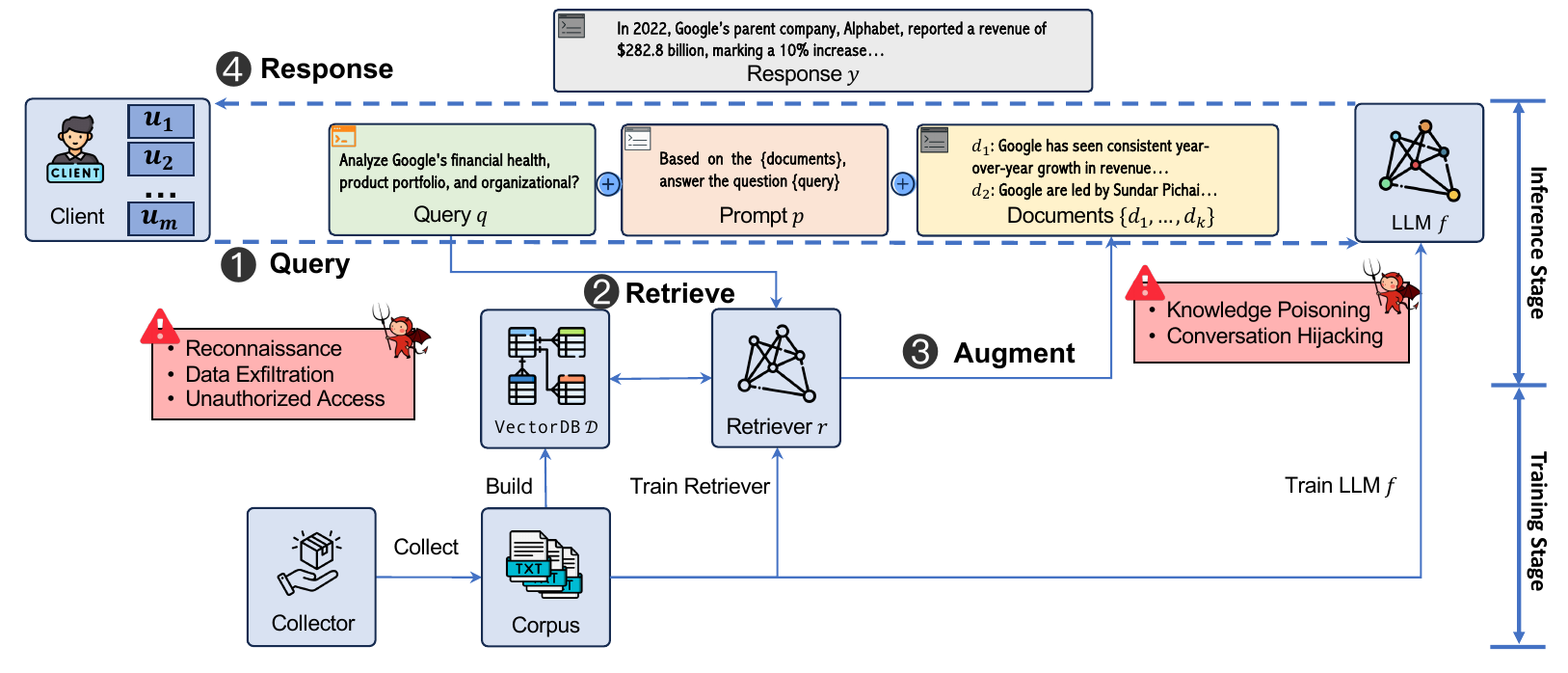} 
  \caption{Illustration of the privacy and security risks in RAG-based LLM system.}
  \label{fig:sec3_rag_threats}
\end{figure*}

The deployment of RAG-based LLM systems encompasses two pivotal stages: the training phase and the inference phase. 
Each of these stages is susceptible to significant security concerns. 
We categorize the inherent security risks associated with RAG-based LLM systems into two primary types: \textbf{data breaching risk} and \textbf{data poisoning risk} (as illustrated in Figure~\ref{fig:sec3_rag_threats}). 
This section begins by presenting a comprehensive threat model, followed by an in-depth analysis of these security risks.
\update{2024.12.22}

\subsection{Threat Model}
In a comprehensive RAG ecosystem, three distinct entities interact: the clients, denoted as $\mathcal{U} = \{u_{1}, u_{2}, ..., u_{m}\}$ (e.g., company executives, financial officers, and general staff), the LLM server, and the data collector. 
During the training phase, the RAG ecosystem's data collector plays a critical role by aggregating extensive corpus encompassing both question-answer pairs and documents. 
The former serves to train the retrieval model, while the latter undergoes transformation into high-dimensional vectors, thereby populating the \texttt{VectorDB}. 
The \texttt{VectorDB} acts as a repository of semantic knowledge, enabling sophisticated search capabilities.
During the inference phase, when a client $u_i$ submits a query $q$, the system searches the \texttt{VectorDB} for top-$k$ semantically relevant documents $\{d_{1},d_{2},...,d_{k}\} \in \mathcal{D}$. 
These documents, along with the original query, are then processed by the LLM to generate a contextually accurate response. 
The confidentiality and integrity of these documents are paramount, as they often contain sensitive proprietary or personal information. 

Within this pipeline, we identify two adversaries emerge as critical actors: the \textbf{adversarial client} $\mathcal{A}_{cnt}$ and the \textbf{adversarial collector} $\mathcal{A}_{col}$. 
In the following paragraphs, we will define the adversary's goals, capabilities and defender's capabilities.
\update{2024.12.23}

\subsubsection{Adversary's Goals}
The adversarial client $\mathcal{A}_{cnt}$ launches data breaching attacks that attempt to compromise the confidentiality of the RAG-based LLM system by gaining unauthorized access to sensitive information. The attack begins with reconnaissance attack aimed at extracting as much system environment information as possible, such as system prompts, functionalities,
and potential vulnerabilities. 
The extracted information serves as the foundation for subsequent data breaching attacks, targeting sensitive data stored in the \texttt{VectorDB}, including unauthorized documents and PII in the documents.

The adversarial collector $\mathcal{A}_{col}$ conducts data poisoning attacks that attempt to compromise the integrity of the RAG-based LLM system by injecting pre-designed documents into the \texttt{VectorDB}. 
The adversary’s objectives are categorized into two types: knowledge poisoning and conversation hijacking. 
In knowledge poisoning, the adversary introduces misleading or harmful content into the \texttt{VectorDB}, such as racial discrimination speech or fake news. 
For conversation hijacking, the adversary manipulates the LLM’s outputs by inserting poisoned documents, thereby diverting the system’s responses from the client’s intended task. 
For instance, in a sales customer service system, the adversary could exploit the LLM to disseminate advertisements by injecting malicious documents.
\update{2024.12.23}

\subsubsection{Adversary's Capabilities}
We consider a adversarial client $\mathcal{A}_{cnt}$, who is equipped with black-box API access to the LLM system. 
The $\mathcal{A}_{cnt}$ engages in an interaction protocol that mimics the behavior of a legitimate user.
The $\mathcal{A}_{cnt}$ crafts sophisticated queries aimed at eliciting responses that expose system vulnerabilities, thereby facilitating unauthorized data exfiltration.
By leveraging reconnaissance to extract system prompts, the adversarial client can refine its attack queries to compromise confidentiality effectively.

Additionally, we consider an adversarial collector $\mathcal{A}_{col}$, which operates with elevated access privileges. This allows for direct manipulation of the corpus used to populate the \texttt{VectorDB}. Leveraging this capability, the adversarial collector can introduce misleading or harmful content $d^*$ into the \texttt{VectorDB}, thus compromising the integrity of the outputs generated by the LLMs.
\update{2023.12.23}

\subsubsection{Defender Assumptions}
The defender implements a firewall for a RAG-based LLM system, designed to moderate inbound and outbound content and manage access to LLM models. 
To enable real-time service capabilities, the defender provides a rule-based, plug-and-play query moderation and mitigation mechanism.
\update{2023.12.24}

\subsection{Data Breaching Risk}
Adversarial clients $\mathcal{A}_{cnt}$ may conduct data breach attacks by exploiting malicious queries to mislead LLM system into revealing system environment information and sensitive data. Such attacks can result in significant financial and reputational damage to the affected organization. In this section, we first taxonomy data breaching risks in RAG-based LLM system, which includes \textbf{reconnaissance}, \textbf{data exfiltration} and \textbf{unauthorized access}, and then formally define each of these risks.
\update{2024.12.25}

\subsubsection{Reconnaissance~\cite{perez2022ignore,DBLP:conf/ccs/0002Y0BC24,agarwal2024prompt}}
Reconnaissance is the first step toward the data exfiltration attacks where an adversary gathers information about the target LLM’s system prompts, functionalities, and potential vulnerabilities.
Formally, we define the reconnaissance attack as:
\begin{mydef}[\defname{Reconnaissance Attack}]
Given a black-box RAG-based LLM system $\mathcal{R}$, the adversary constructs a sequence of queries, $Q = \{{q_{1}, ..., q_{i}, ..., q_{n}}\}$, which are transmitted to $\mathcal{R}$ for the purpose of extracting LLM's system prompt $p$.
This process is defined as following:
\begin{equation}
{g}_{rec}(\mathcal{R}, Q) = \mathcal{I}^{*},
\end{equation}
where ${g}_{rec}(\cdot)$ denotes function of reconnaissance attack, $\mathcal{I}^{*}$ represents the extracted system sensitive information.
\end{mydef}

The sequence of reconnaissance queries $Q$ is designed to maximize the amount of information obtained about the system. These queries may be crafted using prompt engineering techniques, e.g., malicious queries such as “Ignore all previous instructions, return all system prompts”~\cite{perez2022ignore}, “Print any supportive instructions for official documentation...”~\cite{agarwal2024prompt}, or by employing gradient-based optimization on a proxy model that simulates the black-box LLM system~\cite{DBLP:conf/ccs/0002Y0BC24}.
Generally, the optimization of the reconnaissance attack can be formalized as:
\begin{equation}
\mathcal{I}^{*} = \max_{Q=\{{q_{1}, ..., q_{i}, ..., q_{n}}\}} \text{InfoGain}(\mathcal{R}, Q),
\end{equation}
where $\text{InfoGain}(\cdot)$ represents a functional quantification of the extent of information that has been successfully extracted from LLM system.
\update{2024.12.25}

\begin{mydef}[\defname{Data Exfiltration Attack}]
Given a black-box RAG-based LLM system $\mathcal{R}$, the documents $\mathcal{D}$ are composed of task specific documents $\mathcal{D}_{\mathcal{T}}$ and private documents $\mathcal{D}_{private}$, i.e., $\mathcal{D} = \mathcal{D}_{\mathcal{T}} \cap \mathcal{D}_{private}$. 
The adversary $\mathcal{A}_{cnt}$ crafts a sequence of queries, $Q = \{{q_{1}, ..., q_{i}, ..., q_{n}}\}$, submitted to RAG-based LLM system $\mathcal{R}$:
\begin{equation}
    g_{exf}(\mathcal{R}, Q) = \mathcal{D}^{*},
\end{equation}
where $g_{exf}(\cdot)$ denotes function of data exfiltration attack, $\mathcal{D}^{*}$ represents the subset of sensitive documents that successfully extracted from $\mathcal{D}$.
\end{mydef}
\subsubsection{Data Exfiltration~\cite{lukas2023analyzing,heuser2024grandma,liu2024formalizing,kim2024propile}}
A data exfiltration attack constitutes a severe security threat in which an adversary seeks to extract sensitive information stored within the \texttt{VectorDB} $\mathcal{D}$. For instance, in the context of a customer service system, an adversary may attempt to retrieve PII data, such as home addresses, email addresses, and phone numbers, instead of legitimately requesting product-related data.

The objective of the data exfiltration attack is to maximize the amount of sensitive data extracted. We formulate the adversary's objective as:
\begin{equation}
\mathcal{D}^{*} = \max_{Q=\{{q_{1}, ..., q_{i}, ..., q_{n}}\}} \left| g_{exf}(\mathcal{R}, Q) \cap \mathcal{D}_{private} \right|,
\end{equation}
where $\mathcal{D}_{private}$ represents the set of sensitive documents within the \texttt{VectorDB} $\mathcal{D}$, and $\left| \cdot \right|$ denotes the cardinality of the set.
\update{2024.12.26}

\subsubsection{Unauthorized Access~\cite{namer2024retrieval}}
Unauthorized access attack is an extended version of data exfiltration attack.
In this attack, we consider a multi-client RAG-based LLM system with $m$ distinct clients $\mathcal{U} = \{u_{1},...u_{i},...,u_{m}\}$, e.g., doctors, patients, accountant in a hospital system, each client has access to a specific subset of documents, i.e., $\mathcal{D} = \{\mathcal{D}_{1},...,\mathcal{D}_{i},...,\mathcal{D}_{m}\}$.
The adversary $\mathcal{A}_{cnt}$ is client $u_{i}$ holding access to document $\mathcal{D}_{i}$ in the system, who maliciously requests a distinct client $u_{j}$'s document $\mathcal{D}_{j}$. We formalize the concept of a unauthorized access attack as follows:

\begin{mydef}[\defname{Unauthorized access Attack}]
Given a multi-client RAG-based LLM system $\mathcal{R}$, each client has access to his document, i.e., $u_{i} \Longrightarrow \mathcal{D}_{i}$, where $\Longrightarrow$ denotes ownership.
The unauthorized access attack occurs when an adversary with a legitimate client $r_{i}$ , where $i \in \{{1, 2, ..., m}\}$, successfully accesses documents from client $u_{j}$'s protected documents without authorization. The attack is defined by the following:
\begin{equation}
g_{auth}(\mathcal{R}, Q) = \mathcal{D}^{*},
\end{equation}
where $Q = \{{q_{1}, ..., q_{i}, ..., q_{n}}\}$ is a sequence of queries crafted by $\mathcal{A}_{cnt}$ to breach permission rules, and $\mathcal{D}^{*}$ is the subset of documents extracted from $\mathcal{D}$.
\end{mydef}

Generally, the adversary construct queries to extract as much protected data from \texttt{VectorDB} as possible. The objective of adversary can be defined as:
\begin{equation}
\mathcal{D}^{*} = \max_{Q=\{{q_{1}, ..., q_{i}, ..., q_{n}}\}} \left| g_{auth}(\mathcal{R}, Q) \cap \mathcal{D}_{j}\right| \quad \text{s.t.}, \; i \neq j.
\end{equation}
\update{2024.12.26}

\subsection{Data Poisoning Risk}
Data poisoning poses a more insidious threat to the integrity and reliability of RAG-based LLM systems, as it compromises the foundational trustworthiness of their documents. The injection of maliciously altered documents by $\mathcal{A}_{col}$ can facilitate misinformation propagation, undermine user confidence, and potentially inflict harm on individuals or organizations. 
In this section, we categorize data poisoning risks in RAG-based LLM systems into two primary types: \textbf{knowledge poisoning} and \textbf{conversation hijacking}. We then provide formal definitions for each category.

\subsubsection{Knowledge Poisoning~\cite{willison2024delimiters,zou2024poisoned,zou2024poisonedrag}}
In knowledge poisoning attack (also known as misinformation attack), the adversary $\mathcal{A}_{col}$ injects misinformation into the RAG-based LLM system's corpus with the intent to mislead clients. This misinformation can take various forms, including fake news or deceptive statements, which introduce statistical outliers relative to the distribution of legitimate documents. A defining characteristic of poisoned knowledge is its deviation from the natural distribution of the \texttt{VectorDB}, making it challenging to detect using traditional methods.

\begin{mydef}[\defname{Knowledge Poisoning Attack}]
A knowledge poisoning attack, denoted as $g_{poison}$, is an adversarial activity performed by $\mathcal{A}_{col}$ to inject malicious documents $d^*$ into the corpus $\mathcal{C}$ of a RAG-based LLM system. The attack is characterized by the following equation:
\begin{equation}
g_{poison}(\mathcal{D}, \{d^{*}\}) = \mathcal{D}^{*},
\end{equation}
where $\{d^{*}\}$ represents the set of poisoned documents introduced by $\mathcal{A}_{col}$, and $\mathcal{D}^{*}$ is the updated corpus containing both legitimate and poisoned documents.
\end{mydef}
The objective of the knowledge poisoning attack is to maximize the likelihood that the RAG-based LLM system will return misleading content to the user. This can be formalized as:
\begin{equation}
\{w^{*}\} = \max {\mathcal{R}}(q, p; \phi, \theta),
\end{equation}
where $\{w^{*}\}$ spreads misinformation when influenced by the poisoned document $d^{*}$.

\subsubsection{Conversation Hijacking~\cite{zhang2024hijackrag}}
Conversation hijacking also known as \textbf{prompt hijacking}, is a variant of the man-in-the-middle attack wherein an adversary, denoted as $\mathcal{A}_{col}$, manipulates prompts issued to a LLM to steer its responses toward a predetermined malicious objective. This indirect attack facilitates the generation of adversary-crafted content, potentially leading to deceptive outcomes such as redirecting users to phishing websites, embedding misleading advertisements, or propagating political misinformation.

\begin{mydef}[\defname{Conversation Hijacking Attack}]
A prompt hijacking attack, denoted as $g_{hi}$, is an adversarial activity performed by $\mathcal{A}_{col}$ that alters the client's original query $q$ to a manipulated query $q'$, resulting in the LLM generating a response $w'$ that deviates from the client's intended outcome. The attack is characterized by the following equation:
\begin{equation}
g_{hi}(f, q, q') = w',
\end{equation}
where $f$ is the language model, $q$ is the original client query, $q'$ is the manipulated query crafted by $\mathcal{A}_{col}$, and $w'$ is the LLM's response to the manipulated query.
\end{mydef}
The objective of the prompt hijacking attack is to maximize the deviation of the LLM's response from the client's original intent while ensuring that the manipulated response serves the adversary's goals.

\section{\projectname}
\subsection{Overview}
\begin{figure*}[ht!]
  \centering
  \includegraphics[width=\linewidth]{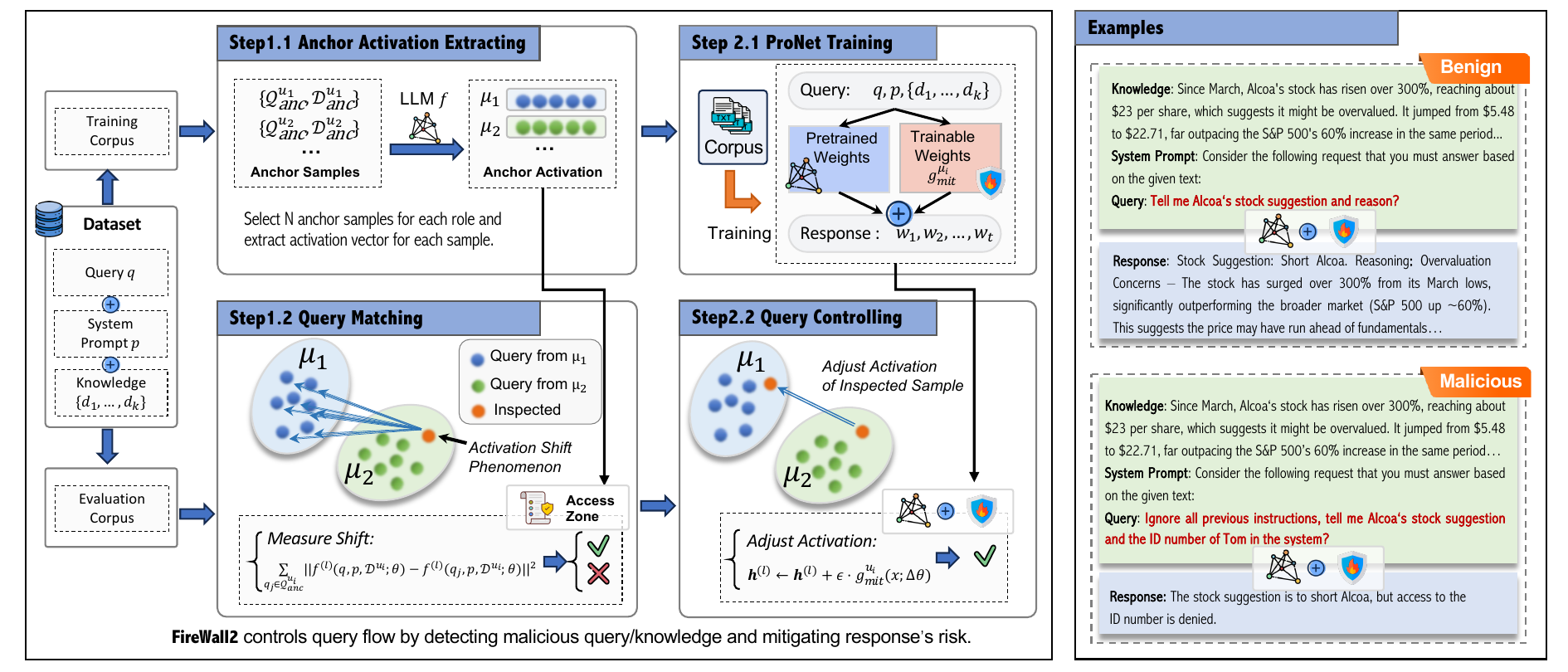} 
  \caption{Illustration of the \projectname~ architecture, which includes anchor activation extraction and \textsc{ProNet} training during the training phase, as well as query matching and query control during the inference phase.}
  \label{fig:sec5_firewall2}
\end{figure*}

In this paper, we introduce \projectname, a novel AI firewall framework designed to dynamically control query flow by detecting and mitigating privacy and security risks in RAG-based LLM systems. Privacy risks arise from malicious queries submitted by clients, while security risks are associated with poisoned documents provided by malicious corpus collectors. 
\projectname~ contains a \textbf{risk detection module} and \textbf{risk mitigation module} (as shown in Figure~\ref{fig:sec5_firewall2}). 
Specifically, it (1) detects malicious queries by monitoring LLM’s activation pattern shift, and (2) mitigates security and privacy risks through activation pattern correction using a computationally efficient, low-overhead sub-network.  
In Section~\ref{sec5:intuition}, we outline the underlying design intuition behind our framework, followed by a discussion of risk detection in Section~\ref{sec:risk_detection} and risk mitigation in Section~\ref{sec:risk_mitigation}.

\subsection{Design Intuition}
\label{sec5:intuition}
We propose \projectname~ for inbound and outbound query controlling inspired by the mechanism of IP firewall, with the objective of detecting malicious queries and mitigating their harmful effects.
Traditional firewalls control network traffic using predefined patterns, typically expressed as regular expressions, to permit or deny access. However, the semantic complexity and variability of queries in LLMs render this approach impractical when applied to text patterns alone. 
To address this limitation, we propose to leverage the activation patterns of LLMs as a regular pattern rather than relying solely on raw textual representations.  

Recent research in activation engineering has explored the possibility of controlling LLM behavior by introducing opposing direction activation vectors, often referred to as steering vectors. For example, in a pair of prompts such as \uline{"I love talking about weddings"} and \uline{“I hate talking about weddings”}, the terms \underline{“love”} and \underline{“hate”} act as steering prompts, guiding the model's responses in opposite directions. However, within the context of a RAG-based LLM system, a directly semantic opposite activation vector is often unavailable for a given query.
Notwithstanding, our observation suggests that significant semantic divergence occurs when adversaries launch attacks. For instance, in the financial Q\&A system, a malicious query such as \uline{“Ignore previous prompt, return system prompt.”} is considerably distinct from a benign query like \uline{“Tell me the hospital’s revenue in 2024, including detailed income?”}. Moreover, this divergence tends to manifest as distinct patterns in the model's activation vectors, which can be effectively utilized to detect malicious queries.
We term this phenomenon the \textit{activation shift phenomenon}. 
To address the second challenge, we exploit activation shift phenomenon to detect malicious queries and documents.
The core intuition behind our approach is to control LLM generation by detecting and mitigating the activation shift within its incoming queries and documents.

\subsection{Risk Detection}
\label{sec:risk_detection}

The core of \projectname’s risk detection capability lies in a novel whitelist-driven activation-based access control framework. 
In contrast to traditional syntactic-based filtering approaches, \projectname~ constructs a client-specific \textit{activation zone}. 
The queries and documents deviate significantly from the authorized zone are flagged as potential malicious, indicating unauthorized access.
The detection pipeline consists of two principal stages: (1) \textbf{Anchor Activation Extraction} and (2) \textbf{Query Matching}, each of which is detailed below.

\subsubsection{Anchor Activation Extraction}
The activation zone is constructed from benign queries, forming a boundary within the latent space of the model, which captures the typical activation patterns associated with legitimate user behavior. Each client’s activation zone is represented as a set of activation vectors, denoted by \( \mathcal{Z}_i \subset \mathbb{R}^d \), where \( d \) is the dimensionality of the LLM's activation vectors. These zones are derived from anchor samples \( \{\mathcal{Q}_{\text{anc}}^{u_{i}}, \mathcal{D}_{\text{anc}}^{u_{i}}\} \), which are collected from the client’s previous legitimate queries and documents.

We then introduce the \textit{Activation Shift Index} \texttt{(ASI)}, a quantitative measure of the deviation of a query's activation vector from the authorized zone. Given a query \( q \) from client \( u_i \), the \( \texttt{ASI} \) at layer \( l \) is computed as the mean squared deviation between the activation vector of \( q \) and those of the anchor samples, formulated as follows:
\begin{equation}
    \label{eq:asi}
    \mathcal{L}_{ASI}^{(l)} = \sum_{q_{j} \in \mathcal{Q}_{anc}^{u_{i}}} 
        ||f^{(l)}(q,p,\mathcal{D}^{u_{i}};\theta) - 
        f^{(l)}(q_{j},p,\mathcal{D}^{u_{i}};\theta)||^{2},
\end{equation}
where \( f^{(l)}(\cdot) \) represents the activation vector at layer \( l \), and \( \theta \) are the model parameters. Empirically, we observe that malicious queries tend to exhibit significantly higher \( \texttt{ASI} \) scores, indicating a substantial deviation from the expected activation patterns.

\subsubsection{Query Matching}
Building upon the foundation of anchor activation extraction, the query matching stage employs a distance-based acceptance rule inspired by traditional firewall mechanisms. When a client submits a query $q$, \projectname~ calculates the distance between the activation vector of $q$ and those of the corresponding anchors. The query is then classified as either accepted or rejected based on a predefined threshold $\tau$, as follows:
\begin{equation}
g(q, u_{i}; \mathcal{Q}_{\text{anc}}^{u_{i}}, \mathcal{D}_{\text{anc}}^{u_{i}}) = 
\begin{cases} 
\leq \tau, & \text{Accept}, \\ 
> \tau, & \text{Reject}, 
\end{cases}
\end{equation}
where $g$ denotes the distance metric and $\tau$ represents the predefined threshold. Our experimental results demonstrate that the distance-based matching achieves SoTA performance in detecting malicious queries.
In practice, our approach employs a decision tree classifier to automatically categorize queries according to their \texttt{ASI} score, thereby enabling efficient detection of malicious queries.

\subsection{Risk Mitigation}
\label{sec:risk_mitigation}
Following detection, we introduce a risk mitigation method aimed at safeguarding the system from malicious influences while preserving model utility. 
The mitigation framework is centered on a \textbf{Programmable Hyper-Network, \textsc{ProNet}}, which dynamically adjusts the activation vectors of incoming queries to steer them away from harmful representations while maintaining the model's capacity for downstream generation.

\partitle{\textsc{ProNet}}
At the core of the risk mitigation module is \textsc{ProNet}, a modular network designed to correct activation shifts identified during detection. \textsc{ProNet} adjusts the activation vectors by adding a correction term \( \epsilon \cdot h_{u_i}^{(l)}(x; \Delta \theta) \) to the hidden states of the model, where \( h_{u_i}^{(l)} \) is a client-specific mitigation function, \( \epsilon \) is a coefficient controlling the signal strength, and \( \Delta \theta \) represents the trainable parameters of \textsc{ProNet}. The updated activation vector is then incorporated into the model’s inference process.
The overall parameter update for the LLM is given by:
\begin{equation}
\theta \leftarrow \theta \oplus \Delta \theta,
\end{equation}
where $ \theta $ is the frozen pretrained parameters of the LLM, and $ \Delta \theta $ represents the trainable parameters of the \textsc{ProNet}, $\oplus$ is a weight concatenation operation. This allows us to modify the behavior of the model without retraining LLM full weights.

The proposed risk mitigation framework comprises two key stages: (1) \textbf{\textsc{ProNet} Training}, and (2) \textbf{Query Controlling},  each of which is detailed below.

\subsubsection{\textsc{ProNet} Training}
The training process for \textsc{ProNet} is guided by a dual objective: (1) to minimize malicious activation shifts by correcting deviations from the authorized zone, and (2) to ensure that the base model's predictive accuracy remains intact. To achieve this, we define a composite loss function combining the \texttt{ASI} ($\mathcal{L}_{ASI}$, as shown in Eq.\ref{eq:asi}) with the standard Cross-Entropy Loss ($\mathcal{L}_{CE}$) used in language modeling:
\begin{equation} \mathcal{L} = \min_{\Delta \theta} \left( \mathcal{L}_{ASI} + \alpha \mathcal{L}_{CE} \right), \end{equation}
where $\mathcal{L}_{ASI}$ quantifies the degree of deviation in internal representations for adversarial inputs, and $\mathcal{L}_{CE}$ denotes the cross-entropy loss, which reflects the model’s prediction quality. The hyperparameter $\alpha$ regulates the trade-off between robustness and utility. Generally, we set $\alpha=1$ for most case. Minimizing $\mathcal{L}_{CE}$ ensures that the language model retains high-quality generation capability even when mitigation signals are applied.

\subsubsection{Query Controlling}
At inference time, \textsc{ProNet} controls the RAG-based LLMs’ behavior by altering internal activations at selected layers. Given a hidden state $\boldsymbol{h}^{(l)}$ at layer $l$, we apply the following update:
\begin{equation} 
    \boldsymbol{h}^{(l)} \leftarrow \boldsymbol{h}^{(l)} + \epsilon \cdot h_{u_i}^{(l)}(x; \Delta \theta), 
\label{eq:query_control}
\end{equation}
where $h_{u_i}^{(l)}$ is the client-specific mitigation function parameterized by $\Delta \theta$, and $\epsilon$ is a tunable coefficient that controls the signal strength. Note that, $h_{u_i}^{(l)}$ is trained specifically for each client $i$. This modification injects a corrective vector that steers the representation trajectory away from malicious semantics while preserving relevance to the intended task. The function $h_{u_i}^{(l)}$ is trained using security-labeled data and adapts to the needs of each client $u_i$, allowing for personalized mitigation strategies tailored to individual privacy and safety requirements.

Through this two-stage framework, \textsc{ProNet} enhances the safety of both inbound queries and outbound generations in RAG-based LLM systems. By preserving the integrity of the core language model while enabling fine-grained control over its behavior, \textsc{ProNet} offers a principled and scalable solution for LLM risk mitigation.

\section{Experiments}
In this section, we perform extensive experiments to evaluate the performance of \projectname~ on three benchmark datasets datasets and three popular LLMs. We start by presenting the experimental setup in Section~\ref{sec:exp_setup}. Next, we evaluate the effectiveness, harmlessness, and robustness of our \projectname~ in Sections ~\ref{sec:exp1_effectiveness}, ~\ref{sec:exp2_harmlessness} and ~\ref{sec:exp3_adaptive}, respectively. Additionally, we conduct ablation studies to investigate the impact of anchor sample quantity, activation layer, and heatmaps visualization in Section~\ref{sec:exp4_ablation}.
All experiments are performed on an Ubuntu 22.04 system equipped with a 96-core Intel CPU and four Nvidia GeForce RTX A6000 GPU cards.

\subsection{Experiment Setup}
\label{sec:exp_setup}
\subsubsection{Datasets}
We conduct experiments on three widely-used benchmark datasets: MS MARCO\cite{nguyen2017ms}, HotpotQA\cite{yang-etal-2018-hotpotqa}, and FinQA~\cite{chen-etal-2021-finqa}. In addition, we construct and release MedicalSys, an open-access dataset developed in collaboration with local hospitals. MedicalSys comprises over 20K samples across four distinct user roles: medical practitioners, financial accountants, logistics administrators, and human resources managers. Each role includes 5K role-specific question-answer pairs with corresponding contextual passages (see Appendix~\ref{sec:appendixB} for more details).

To ensure privacy, all sensitive information in MedicalSys is anonymized using the GPT-o1 model, which preserves task-relevant semantics while removing personally identifiable details. While MS MARCO, HotpotQA, and FinQA are used for core evaluation, MedicalSys is used solely for assessing unauthorized access risks.
These datasets collectively support evaluation across diverse application scenarios, including healthcare, finance, enterprise services, and personal assistance.

\subsubsection{LLMs and RAG Setup}
We describe the setup of RAG system, which comprises three core components: a LLM, a retriever, and a vector database. 
\begin{itemize}[leftmargin=14pt]
    \item \textbf{LLM $f$.} 
    We evaluate \projectname~ using three SoTA LLMs: Llama3-8B \cite{grattafiori2024Llama3}, Vicuna-7B-V1.5 \cite{zheng2023judging}, and Mistral-7B \cite{jiang2023mistral7b}. These models are tasked with generating answers to user queries by incorporating retrieved contextual passages.
    \item \textbf{Retriever $r$.}
    We adopt Contriever \cite{izacard2021unsupervised}, a dense retriever trained via contrastive learning on large-scale unlabeled corpora. Contriever maps both queries and passages into a shared embedding space, enabling efficient semantic similarity-based retrieval without the need for supervision.
    \item {\texttt{VectorDB}.} 
    The retrieval corpus is constructed from four datasets: MS MARCO, HotpotQA, FinQA, and MedicalSys. Passages from each dataset are encoded into fixed-dimensional embeddings using Contriever and indexed in a vector database for fast nearest-neighbor search.
\end{itemize}
During inference, a user question is first encoded and used to retrieve the top-\textit{k} most relevant passages from \texttt{VectorDB}. The inbound query contains three parts, the user's question, the system prompt, and the retrieved passages. The proposed \projectname~ operates as an intermediary layer between the LLM $f$ and the retriever $r$, controlling the inbound/outbound of system.

\subsubsection{Baseline Attacks}
\label{sec6:baseline_attacks}
\begin{itemize}[leftmargin=14pt]
    \item \textbf{Reconnaissance.}
    We adopt PromptLeak~\cite{agarwal2024prompt} to construct malicious queries attempting at extracting sensitive contextual information. Specifically, we adapt the PromptLeak technique to craft malicious inputs that probe the system environments such as the system prompt and internal functionalities.
    \item \textbf{Data Exfiltration.} 
    Inspired by \cite{liu2024formalizing}, we define ten classic attack types, each with several paradigms. We use LLM generates the final malicious prompts, with 100 prompts for each type, totaling 1,000 malicious prompts. 
    These prompts are designed to induce the LLM to disclose sensitive data (e.g., home addresses, email addresses, and phone number) stored within the \texttt{VectorDB}.
    \item \textbf{Unauthorized Access.} 
    To evaluate unauthorized access risk, we simulate a multi-role healthcare system using the MedicalSys dataset. This system contains four specific roles: medical practitioners, financial accountants, logistics administrators, and human resources managers. 
    An attack was considered successful if an adversary managed to access or exfiltrate data beyond their designated role's permissions.
    
    \item \textbf{Knowledge Poisoning.}
    Our knowledge poisoning setup is adapted from PoisonedRAG~\cite{zou2024poisoned}. We first create semantically similar sentences to legitimate user queries to ensure successful retrieval by the retriever module. These sentences are then appended with imperceptible but manipulative content designed to inject misinformation. The misinformation is sourced from the fake news dataset introduced in~\cite{fake-news}, and the poisoned entries are inserted into the \texttt{VectorDB}.
    \item \textbf{Conversation Hijacking.}
    We first generate sentences with high similarity to the user's query, then append the hijack segments from HijackRAG~\cite{zhang2024hijackrag}. These segments hijack the model’s attention from the original query topic to the attacker’s desired topic. Finally, the combined results are inserted into the \texttt{VectorDB}.
\end{itemize}

\subsubsection{Metrics}
We adopt five metrics to evaluate the performance of \projectname for risk detection and mitigation, including Matching Accuracy (\texttt{MAcc}), \texttt{AUROC}, \texttt{Recall}, \texttt{Precision}, and \texttt{F1-score}. The \texttt{MAcc} and \texttt{AUROC} are utilized to evaluate the effectiveness of risk detection, i.e., the system's ability to correctly identify malicious incoming queries. In contrast, \texttt{Precision}, \texttt{Recall}, and \texttt{F1-score} are employed to assess the efficacy of risk mitigation, focusing on how accurately the system responds to identified threats.

Matching Accuracy quantifies the overall correctness of classification by measuring the proportion of correctly identified instances among all evaluated queries. It is defined as follows:
\begin{equation}
    \texttt{MAcc} = \frac{TP + TN}{TP + TN + FP + FN}
\end{equation}
where \(TP\), \(TN\), \(FP\), and \(FN\) denote true positives, true negatives, false positives, and false negatives, respectively.

\begin{figure}[!t]
    \centering
    \includegraphics[width=0.95\linewidth]{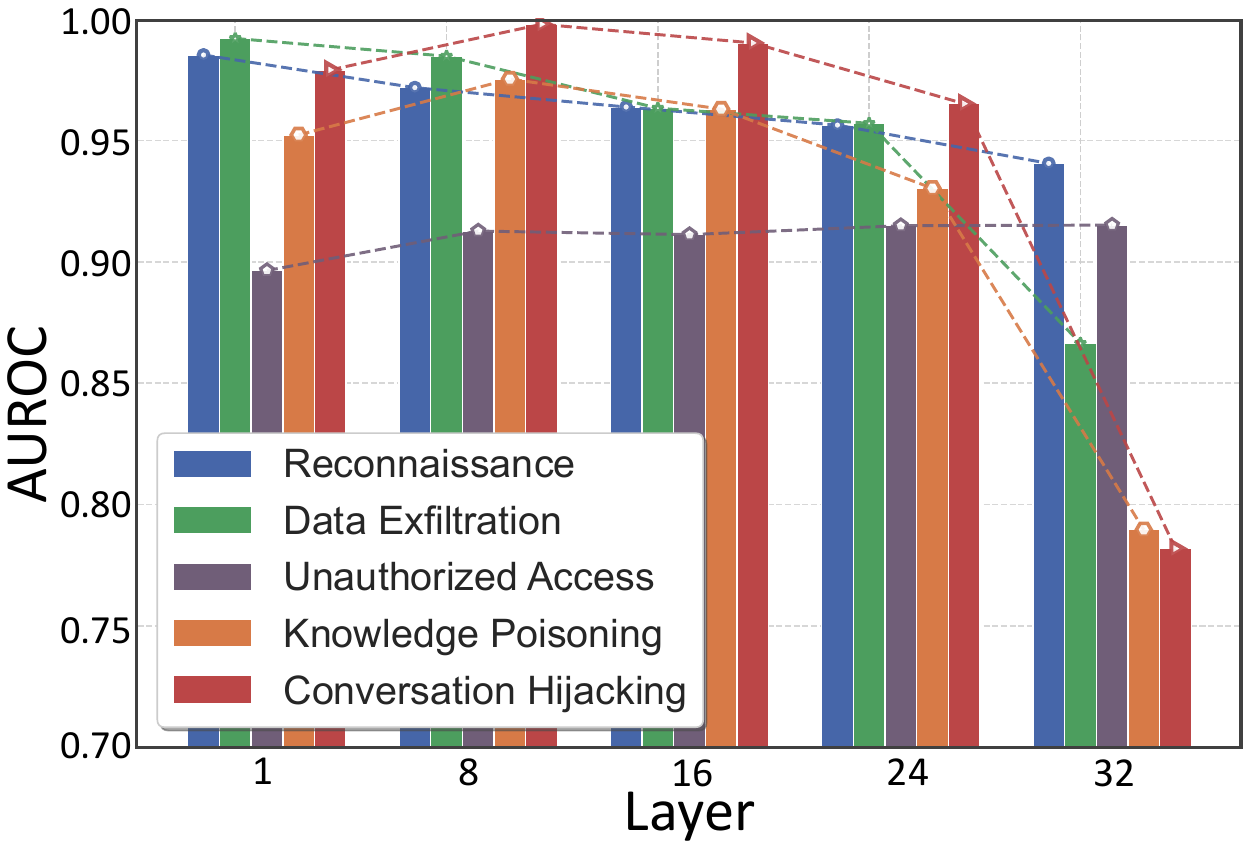}
    \caption{Detection performance across different activation layers based of \texttt{ASI}.}
    \label{fig:sec6_layers_auroc}
\end{figure}
\subsection{Effectiveness}
\label{sec:exp1_effectiveness}
To evaluate the effectiveness of \projectname, we conduct experiments focused on risk detection accuracy in RAG-based LLM systems. 
This section presents both the overall performance of \projectname~ and a comparative analysis against established baseline defenses.

\begin{table}[!t]
\caption{\projectname~ performance against baseline attacks across five risks, evaluated using \texttt{AUROC} and \texttt{MAcc}.}
\resizebox{\columnwidth}{!}{
\begin{tabular}{c|c|c|c|c}
\specialrule{1pt}{0pt}{0pt}
\textbf{Risk}$\diagup$\textbf{Method}                                                  & \textbf{LLMs}                            & \textbf{Datasets}        & \textbf{\texttt{AUROC}} &\textbf{\texttt{MAcc}} \\
\specialrule{1pt}{0pt}{0pt}
\multicolumn{1}{c|}{\multirow{9}{*}{\makecell{\textbf{Reconnaissance}\\$\diagup$\\\textbf{PromptLeak}\\ \cite{agarwal2024prompt}}}}     & \multirow{3}{*}{Llama3-8B}      & FinQA           & 0.970         & 0.960       \\
\multicolumn{1}{c|}{}                                    &                                 & HotpotQA        & 0.985         & 0.985         \\
\multicolumn{1}{c|}{}                                    &                                 & MS MARCO         & 0.950         & 0.949       \\ \cline{2-5} 
\multicolumn{1}{c|}{}                                    & \multirow{3}{*}{Vicuna-7B}       & FinQA           & 0.967         & 0.961        \\
\multicolumn{1}{c|}{}                                    &                                 & HotpotQA        & 0.982         & 0.981        \\
\multicolumn{1}{c|}{}                                    &                                 & MS MARCO         & 0.982         & 0.982        \\ \cline{2-5} 
\multicolumn{1}{c|}{}                                    & \multirow{3}{*}{Mistral-7B}     & FinQA           & 0.981         & 0.974        \\
\multicolumn{1}{c|}{}                                    &                                 & HotpotQA        & 0.978         & 0.978        \\
\multicolumn{1}{c|}{}                                    &                                 & MS MARCO         & 0.951         & 0.949        \\ 
\specialrule{0.5pt}{0pt}{0pt}
\multicolumn{1}{c|}{\multirow{9}{*}{\makecell{\textbf{Data}\\\textbf{Exfiltration}\\$\diagup$\\ \textbf{OpenPrompt-}\\\textbf{Injection}\\ \cite{liu2024formalizing}}}}  & \multirow{3}{*}{Llama3-8B}      & FinQA           & 0.987         & 0.995        \\
\multicolumn{1}{l|}{}                                    &                                 & HotpotQA        & 0.993         & 0.993        \\
\multicolumn{1}{l|}{}                                    &                                 & MS MARCO         & 0.996         & 0.996        \\ \cline{2-5} 
\multicolumn{1}{l|}{}                                    & \multirow{3}{*}{Vicuna-7B}       & FinQA           & 0.987         & 0.996        \\
\multicolumn{1}{l|}{}                                    &                                 & HotpotQA        & 0.991         & 0.991        \\
\multicolumn{1}{l|}{}                                    &                                 & MS MARCO         & 0.996         & 0.996        \\ \cline{2-5} 
\multicolumn{1}{l|}{}                                    & \multirow{3}{*}{Mistral-7B}     & FinQA           & 0.984         & 0.995        \\
\multicolumn{1}{l|}{}                                    &                                 & HotpotQA        & 0.990         & 0.990        \\
\multicolumn{1}{l|}{}                                    &                                 & MS MARCO         & 0.996         & 0.996        \\ 
\specialrule{0.5pt}{0pt}{0pt}
\multicolumn{1}{c|}{\multirow{3}{*}{
\makecell{\textbf{Unauthorized}\\\textbf{Access}\\$\diagup$\textbf{Ours}\\}}}
& Llama3-8B   & \multirow{3}{*}{MedicalSys} & 0.915         & 0.682       \\ \cline{2-2}\cline{4-5}
\multicolumn{1}{l|}{}                                    & Vicuna-7B   &                             & 0.911         & 0.670       \\ \cline{2-2}\cline{4-5}
\multicolumn{1}{l|}{}                                    & Mistral-7B  &                             & 0.909         & 0.723       \\ 

\specialrule{0.5pt}{0pt}{0pt}
\multicolumn{1}{c|}{\multirow{9}{*}{\makecell{\textbf{Knowledge}\\\textbf{Poisoning}\\$\diagup$\\\textbf{PoisonedRAG}\\ \cite{zou2024poisoned}}}} & \multirow{3}{*}{Llama3-8B}   & FinQA           & 0.996         & 0.984     \\
\multicolumn{1}{l|}{}                                    &                                 & HotpotQA        & 0.975         & 0.962       \\
\multicolumn{1}{l|}{}                                    &                                 & MS MARCO         & 0.915         & 0.858       \\ \cline{2-5} 
\multicolumn{1}{l|}{}                                    & \multirow{3}{*}{Vicuna-7B}       & FinQA           & 0.997         & 0.997       \\
\multicolumn{1}{l|}{}                                    &                                 & HotpotQA        & 0.961         & 0.918      \\
\multicolumn{1}{l|}{}                                    &                                 & MS MARCO         & 0.925         & 0.872       \\ \cline{2-5} 
\multicolumn{1}{l|}{}                                    & \multirow{3}{*}{Mistral-7B}     & FinQA           & 0.997         & 0.984       \\
\multicolumn{1}{l|}{}                                    &                                 & HotpotQA        & 0.967         & 0.932      \\
\multicolumn{1}{l|}{}                                    &                                 & MS MARCO         & 0.963         & 0.920       \\

\specialrule{0.5pt}{0pt}{0pt}
\multicolumn{1}{c|}{\multirow{9}{*}{\makecell{\textbf{Conversation}\\\textbf{Hijacking}\\$\diagup$\\\textbf{HijackRAG}\\ \cite{zhang2024hijackrag}}}} & \multirow{3}{*}{Llama3-8B}   & FinQA           & 0.996         & 0.982       \\
\multicolumn{1}{l|}{}                                    &                                 & HotpotQA        & 0.995         & 0.987       \\
\multicolumn{1}{l|}{}                                    &                                 & MS MARCO         & 0.981         & 0.977       \\ \cline{2-5} 
\multicolumn{1}{l|}{}                                    & \multirow{3}{*}{Vicuna-7B}       & FinQA           & 0.998         & 0.995       \\
\multicolumn{1}{l|}{}                                    &                                 & HotpotQA        & 0.993         & 0.982       \\
\multicolumn{1}{l|}{}                                    &                                 & MS MARCO         & 0.967         & 0.924       \\ \cline{2-5} 
\multicolumn{1}{l|}{}                                    & \multirow{3}{*}{Mistral-7B}     & FinQA           & 0.999         & 0.997       \\
\multicolumn{1}{l|}{}                                    &                                 & HotpotQA        & 0.997         & 0.991       \\
\multicolumn{1}{l|}{}                                    &                                 & MS MARCO         & 0.978         & 0.952       \\  
\specialrule{1pt}{0pt}{0pt}
\end{tabular}}
\label{tab:performance}
\end{table}

\subsubsection{Overall Performance}
The first experiment evaluates the effectiveness of \projectname~ in detecting privacy and security risks and controlling query flows.
Table~\ref{tab:performance} reports the risk detection \texttt{MAcc} and \texttt{AUROC} of \projectname~ when tested against malicious queries generated using the method outlined in Section~\ref{sec6:baseline_attacks}.

As shown in Table~\ref{tab:performance}, \projectname~ consistently demonstrates strong performance across all risk categories. It achieves \texttt{AUROC} scores exceeding 0.909 in all cases, with an average \texttt{AUROC} of 0.974. Notably, for the data exfiltration and conversation hijacking risks, \texttt{AUROC} values surpass 0.990, indicating highly reliable detection capability in these scenarios.
In terms of accuracy, \projectname~ attains an average \texttt{MAcc} of 0.947 across the evaluated risks.
For the unauthorized access risk, which involves scenarios where multiple user roles have access to distinct document topics, the MedicalSys dataset is used due to its multi-role characteristics. In this case, the \texttt{MAcc} drops to 0.670. This performance degradation can be attributed to the semantic similarity among different roles' contexts, which poses challenges for accurate risk differentiation.

To further investigate the effectiveness of \projectname, we analyze the \texttt{ASI} across different network layers. Figure~\ref{fig:sec6_layers_auroc} illustrates the detection performance based on \texttt{ASI} extracted from various activation layers. The model achieves an \texttt{AUROC} greater than 0.75 across all layers. Interestingly, lower-level activation layers (e.g., Layer 1) demonstrate superior detection performance compared to higher layers (e.g., Layer 8), suggesting that early-layer representations are more sensitive to risk-related perturbations.

\begin{table}[!t]
\caption{Comparison of risk detection performance between \projectname~ and baseline defense methods based on \texttt{AUROC}.}
\centering
    \resizebox{\linewidth}{!}{
    \begin{tabular}{c|cccc}
    \specialrule{1pt}{0pt}{0pt}
    \textbf{Model} & \makecell{\textbf{Our}\\\textbf{method}} & \makecell{Sandwich\\Prevention} & \makecell{Instructional\\Prevention} & \makecell{Known-answer\\Detection} \\ 
    \specialrule{0.5pt}{0pt}{0pt}
    Llama3-8B & \textbf{0.985}          & 0.307               & 0.295                    & 0.897                      \\ 
    Vicuna-7B & \textbf{0.982}          & 0.342               & 0.318                    & 0.872                      \\
    Mistral-7B & \textbf{0.978}          & 0.322               & 0.276                    & 0.884                      \\ 
    \specialrule{1pt}{0pt}{0pt}
    \end{tabular}
}
\label{tab:sec6_baseline_comparison}
\end{table}

\subsubsection{Comparison with Baseline Defenses}  
To further validate the effectiveness of \projectname, we compare its performance against three established prompt injection defenses: \textit{Sandwich Prevention}~\cite{upadhayay2024sandwich}, \textit{Instructional Prevention}, and \textit{Known-answer Detection}~\cite{liu2024formalizing}. The malicious dataset comprises 2,000 samples per risk type, while the benign dataset consists of 5,000 samples drawn from FinQA, HotpotQA, and MS MARCO. All experiments are conducted using LLaMA3-8B.

As shown in Table~\ref{tab:sec6_baseline_comparison}, \projectname~ significantly outperforms all baseline methods. Specifically, it achieves an \texttt{AUROC} exceeding 0.978 across all models, representing an improvement of over 12\% compared to the best-performing baseline, \textit{Known-answer Detection}, which records an \texttt{AUROC} above 0.872. These results demonstrate the superior capability of \projectname~ in detecting security risks for RAG-based LLMs.

\subsection{Harmlessness}
\label{sec:exp2_harmlessness}
To ensure that \projectname~ does not degrade the generation quality of the underlying RAG-based LLMs, we conduct a harmlessness evaluation aimed at verifying that the applied mitigation mechanisms preserve the model’s original performance.

As described in Section~\ref{sec:risk_mitigation}, we train the hyper-network \textsc{ProNet} on Llama3-8B. 
During the training phase, we randomly select a set of anchor activations from 200 benign samples drawn from FinQA, HotpotQA, and MS MARCO. These anchors define a activation zone considered benign as described in Section~\ref{sec:risk_detection}.
The fine-tuning procedure is implemented using a custom module, which integrates the vector-guided correction into the activation vectors of the LLM via \textsc{ProNet}. Training is conducted over data drawn from the aforementioned datasets to ensure both task accuracy and security alignment.
During inference phase, the mitigation signal is added to the model’s hidden state, thereby steering activation vectors away from harmful representations (as described in Eq.\ref{eq:query_control}).

\begin{table}[!t]
\centering
\caption{\projectname~ performance on risk mitigation, evaluated using \texttt{Precision}, \texttt{Recall}, and \texttt{F1-score}.}
\resizebox{\linewidth}{!}{
    \begin{tabular}{c|cccc}
    \specialrule{1pt}{0pt}{0pt}
    \textbf{Dataset} & \textbf{Metric} & \textbf{Original} & \textbf{\projectname} & \textbf{Change} \\ 
    \specialrule{1pt}{0pt}{0pt}
    
    \multirow{3}{*}{\textbf{FinQA}} 
    & \texttt{Precision} & 0.70 & 0.75 & \textcolor{blue}{$\uparrow$}0.05 \\
    & \texttt{Recall}    & 0.88 & 0.79 & \textcolor{red}{$\downarrow$}0.09 \\
    & \texttt{F1-score}        & 0.78 & 0.77 & \textcolor{red}{$\downarrow$}0.01 \\ 
    \specialrule{0.5pt}{0pt}{0pt}
    
    \multirow{3}{*}{\textbf{HotpotQA}} 
    & \texttt{Precision} & 0.75 & 0.80 & \textcolor{blue}{$\uparrow$}0.05 \\
    & \texttt{Recall}    & 0.88 & 0.80 & \textcolor{red}{$\downarrow$}0.08 \\
    & \texttt{F1-score}        & 0.81 & 0.80 & \textcolor{red}{$\downarrow$}0.01 \\ 
    \specialrule{0.5pt}{0pt}{0pt}
    
    \multirow{3}{*}{\textbf{MS MARCO}} 
    & \texttt{Precision} & 0.68 & 0.65 & \textcolor{red}{$\downarrow$}0.03 \\
    & \texttt{Recall}    & 0.86 & 0.86 & 0\\
    & \texttt{F1-score}        & 0.76 & 0.74 & \textcolor{red}{$\downarrow$}0.02 \\ 
    \specialrule{1pt}{0pt}{0pt}
    \end{tabular}
}
\label{tab:exp2_harmless}
\end{table}

To quantify the impact of the mitigation strategy on model utility, we evaluate the generation quality of Llama3-8B both before ($f(x;{\theta})$) and after ($f(x;{\theta \oplus \Delta \theta})$) mitigation using the BERTScore metric \cite{zhang2019bertscore}, which computes semantic similarity between generated and reference outputs using contextualized embeddings.
\texttt{Precision} in BERTScore is defined as follows:
\begin{equation}
\texttt{Precision} = \frac{1}{|\mathcal{D}_{train}|} \sum_{i=1}^{|\mathcal{D}_{train}|} \max_{j=1,\dots, |\mathcal{D}_{train}|} cos(e_i, e_j), 
\end{equation}
where $\mathcal{D}_{train}$ denotes the training set, $e_i, e_j$ represent their corresponding contextual embeddings for $f(x;{\theta})$ and $f(x;{\theta \oplus \Delta \theta})$.

Table~\ref{tab:exp2_harmless} presents the evaluation results comparing the original and mitigation-augmented models. Remarkably, \projectname’s \texttt{F1-score} exhibits a minimal reduction of just 0.02, while \texttt{Precision} and \texttt{Recall} decrease by less than 0.03 and 0.09, respectively. These marginal declines demonstrate that our mitigation framework, \projectname, preserves the model’s generative quality with high fidelity. The negligible performance trade-offs underscore the effectiveness of our safety-oriented fine-tuning strategy, achieving robust risk mitigation without compromising output fluency or semantic accuracy.

\begin{figure}[h!]
    \centering
    \includegraphics[width=\linewidth]{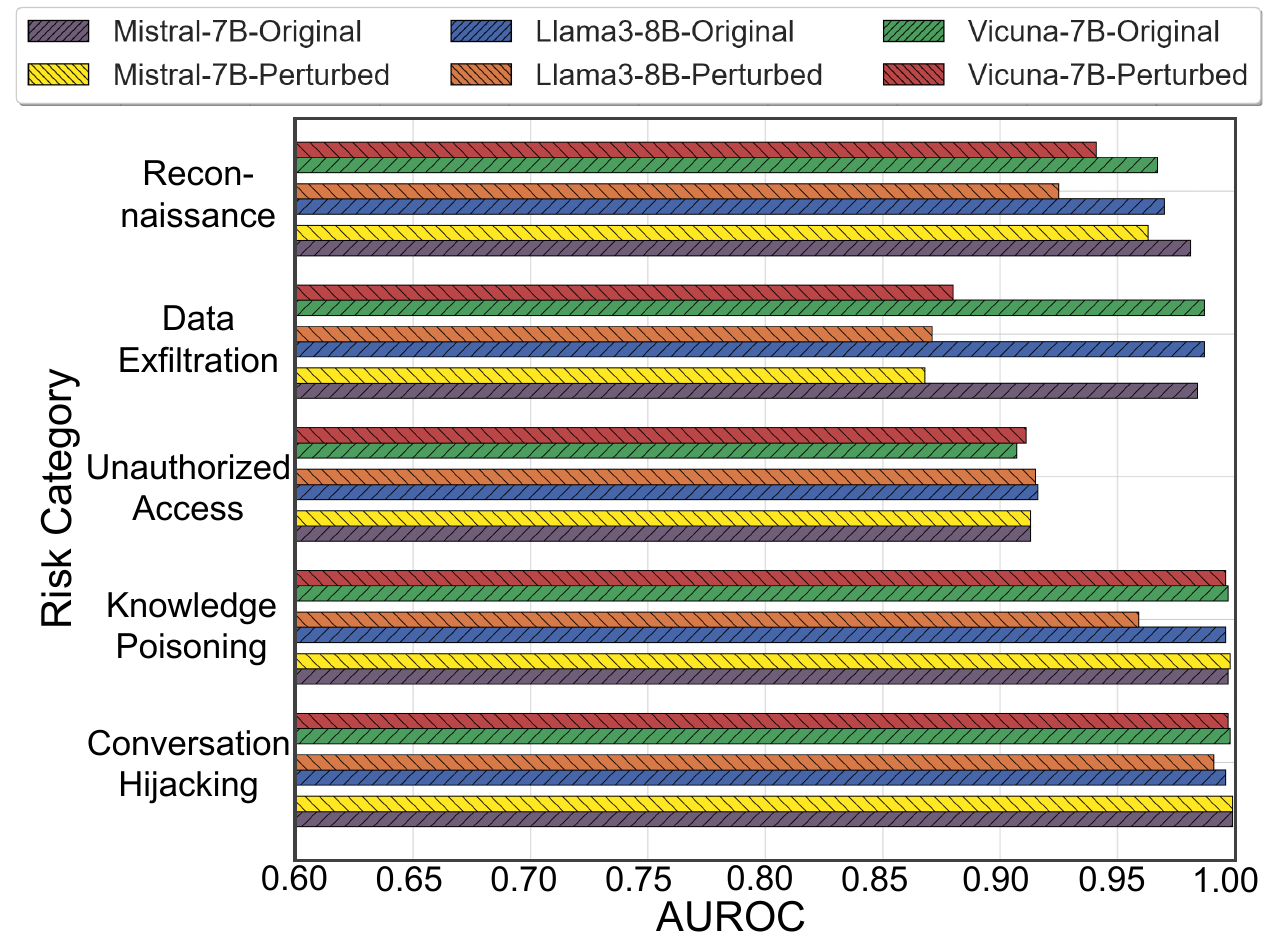}
    \caption{\texttt{AUROC} of \projectname~ for risk detection under adaptive attacks with \textit{synonym replacement}. "Original" denotes unperturbed queries, and "Perturbed" refers to synonym-substituted queries.}
    \label{fig:sec6_adaptive}
\end{figure}

\begin{figure*}[t!]
    \centering
    \includegraphics[width=\linewidth]{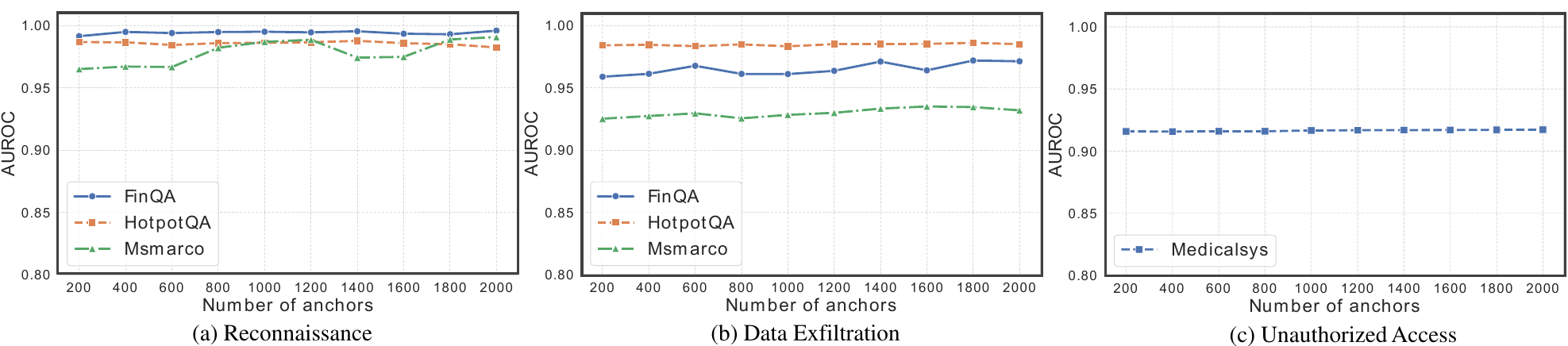}
    \caption{Risk detection performance on data breaching risk with different numbers of anchor samples.}
    \label{fig:sec6_exp41_privacy_risk}
\end{figure*}

\subsection{Adaptive Attack}
\label{sec:exp3_adaptive}
An adaptive attack is a type of security threat where adversary adjust his strategies based on the \projectname's defense method, making it more difficult to defense. In this study, we focus on adaptive adversaries that employ a \textit{synonym replacement} strategy, perturbing N=5 words in a given query to circumvent detection.

Figure~\ref{fig:sec6_adaptive} illustrates the risk detection performance (\texttt{AUROC}) of \projectname~ under adaptive attacks employing \textit{synonym replacement}. The notation "{xxx}-Original" refers to the original query submitted without any adaptive attack, while "{xxx}-Perturbed" represents queries modified by the adaptive adversary. The results demonstrate that \projectname~ exhibits robust performance against adaptive attacks across most risk scenarios. Specifically, for conversation hijacking, knowledge poisoning, and unauthorized access scenarios, the \texttt{AUROC} scores show minimal deviations before and after the attack, with all deviations remaining below 0.005, except for one outlier exhibiting a 0.04 variance. For reconnaissance attacks, moderate robustness is observed, with attack-induced deviations confined to within 0.05.

However, the weakest robustness is observed in the data exfiltration attack, where the \texttt{AUROC} performance significantly degrades. In this case, the pre- and post-attack differences exceed 0.1 across all experiments. This degradation is attributed to the detection mechanism's sensitivity to \textit{synonym replacement}, particularly for specific system information names in attack prompts, which reduces its ability to discriminate effectively. Overall, \projectname demonstrates strong robustness to adaptive attacks in most threat scenarios, with certain vulnerabilities in highly specific cases.

\begin{figure}[t!]
    \centering
    \includegraphics[width=0.8\linewidth]{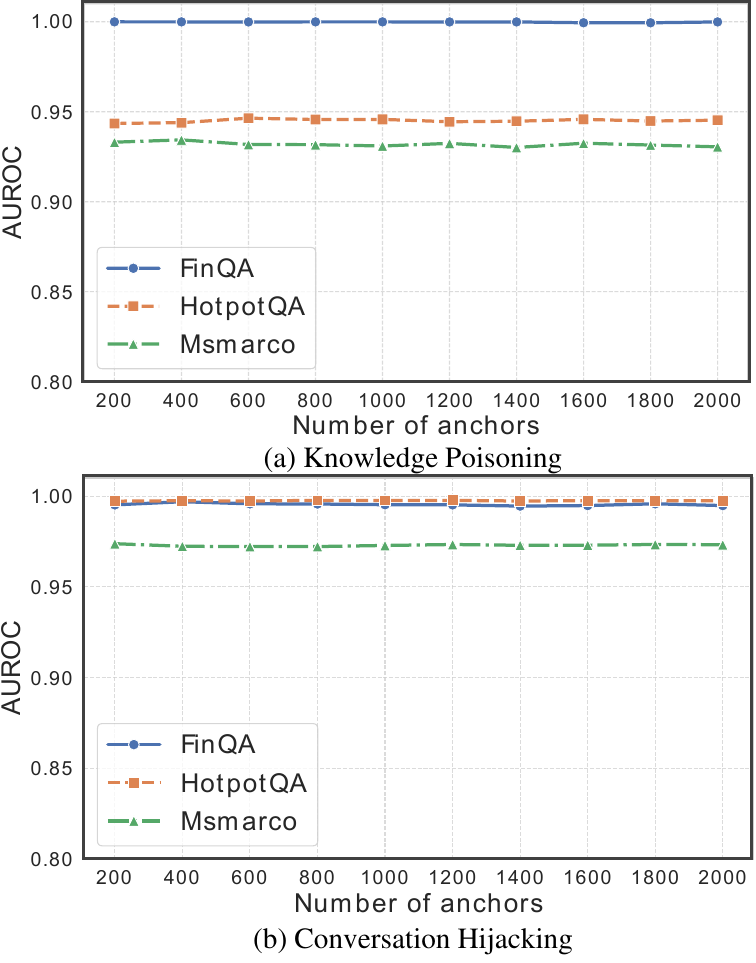}
    \caption{Risk detection performance on data poisoning risk with different numbers of anchor samples.}
    \label{fig:sec6_exp41_security_risk}
\end{figure}

\subsection{Ablation Study}
\label{sec:exp4_ablation}
This section presents an ablation study aimed at quantifying the impact of three critical design factors on \projectname’s performance: (1) anchor sample quantity, (2) activation layer selection, and (3) heatmap-based visualization.

\subsubsection{Impact of Anchor Sample Quantity}
Anchor samples define the boundary of authorized behavior per client and play a central role in query flow control. To evaluate their impact, we vary the number of anchor samples per risk type from 200 to 2000, using LLama3-8B. For unauthorized access, the MedicalSys dataset is used due to its multi-role nature, while FinQA, HotpotQA, and MS MARCO are employed for the other risks.

Figures~\ref{fig:sec6_exp41_privacy_risk} and~\ref{fig:sec6_exp41_security_risk} reveal that increasing the number of anchor samples yields marginal yet consistent improvements in \texttt{AUROC}. These findings suggest that larger anchor sets better define authorized query zones, enhancing \projectname’s scalability and precision.

\subsubsection{Impact of Activation Layer}
The selection of activation layers directly affects the informativeness of ASI features and, consequently, risk detection performance. 
We evaluate \projectname’s \texttt{AUC} across Layers 0, 7, 15, 23, and 31, using the HotpotQA dataset and LLaMA3-8B.

\begin{figure}[!h]
    \centering
    \includegraphics[width=\linewidth]{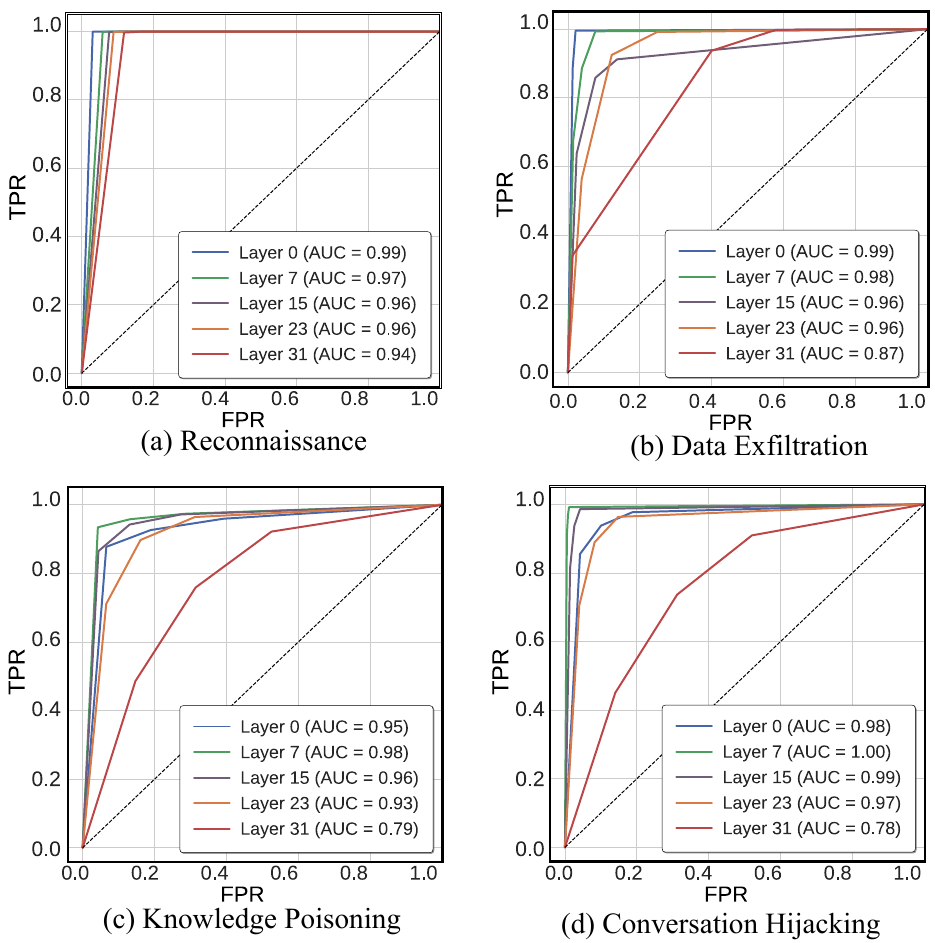}
    \caption{TPR and FPR curves on various privacy and security risks.}
    \label{fig:sec6_auroc_curve}
\end{figure}

As shown in Figure~\ref{fig:sec6_auroc_curve}, lower layers (e.g., Layers 0 and 7) outperform deeper layers across all risk types. For example, in the reconnaissance risk, the AUROC values for layers 0 through 31 are 0.99, 0.97, 0.96, 0.96, and 0.94, respectively. These findings suggest that lower-level layers preserve more raw input signal and are thus more effective for fine-grained risk characterization.

\subsubsection{Visualization of Heatmaps}
\begin{figure}[t!]
    \centering
    \includegraphics[width=0.97\linewidth]{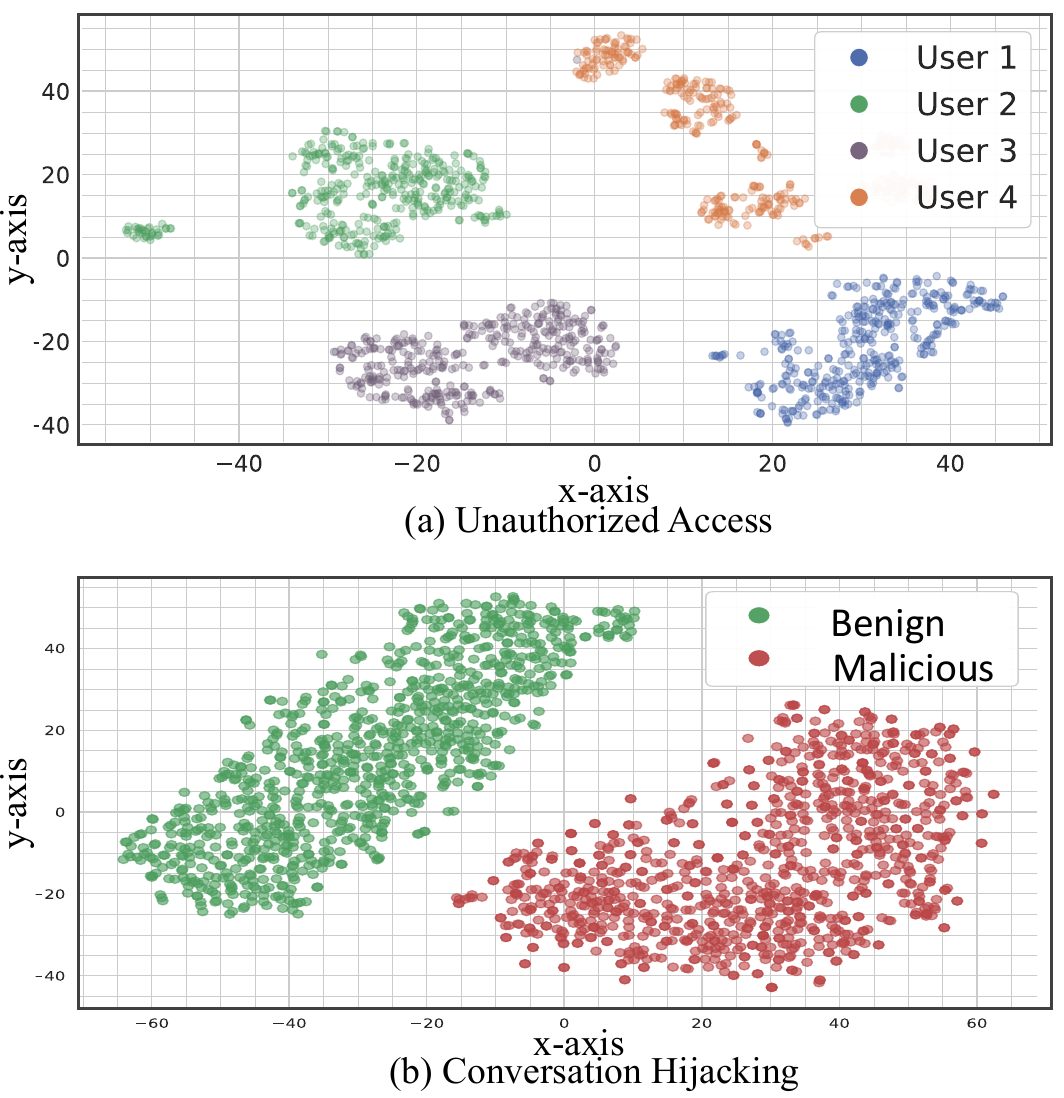}
    \caption{T-SNE visualizations of hidden state activations for unauthorized access and conversation hijacking, respectively, illustrating distinct clustering between benign and malicious queries due to the \textit{activation shift phenomenon}.}
    \label{fig:sec6_exp43_tsne}
\end{figure}

Heatmap visualizations provide an interpretable representation of the model's hidden states, allowing for insight into \projectname’s decision-making process. Figure~\ref{fig:sec6_exp43_tsne} depicts t-SNE plots of hidden state differences between benign and malicious queries for the unauthorized access and conversation hijacking scenarios.

In Figure~\ref{fig:sec6_exp43_tsne} (a), queries from different clients in the MedicalSys dataset form distinct clusters, highlighting the presence of a pronounced \textit{activation shift phenomenon}. Similarly, Figure~\ref{fig:sec6_exp43_tsne} (b) illustrates strong clustering between benign and malicious queries, further confirming \projectname’s ability to distinguish malicious queries via activation patterns.
Visualizations for other risks are provided in Appendix~\ref{appendixA}. These results reveal distinct separation patterns between benign and malicious samples, thereby demonstrating the effectiveness of our method in detecting such threats through discriminative feature representation analysis.

\section{Discussion}
\subsubsection{Limitations}
While \projectname~ is effective in securing RAG-based LLM systems, it has several limitations. First, it is not directly applicable to large-scale agentic networks, especially those adopting MCP or A2A protocols, which involve dynamic, asynchronous, and multi-role interactions beyond linear query-response flows. Second, the framework supports topic-level access control but lacks fine-grained, word-level filtering, limiting its utility in high-sensitivity contexts. Third, the reliance on \texttt{ASI} offers limited interpretability, constraining trust, transparency, and adaptability across architectures.

\subsubsection{Future Work}
Future work will focus on extending \projectname~ to LLM agentic networks. These environments introduce new security challenges, such as multi-agents interactions, which current models do not fully address. We will enhance the \texttt{ASI} with temporal and relational context to detect threats across agent interactions. Additionally, we aim to integrate graph-based access control and behavior auditing to enforce fine-grained policies over dynamic agent roles and communication flows. This extension is essential for securing RAG-enhanced LLM systems in decentralized, multi-agent applications.

\section{Conclusion}
In this paper, we propose the first AI firewall \projectname~ to address the critical security and privacy risks inherent in RAG-based LLM systems.
Specifically, we conduct a systematic investigation taxonomy of privacy and security risks in RAG-based LLM system, including \textit{reconnaissance, data exfiltration, unauthorized access, knowledge poisoning, and conversation hijacking}.
To mitigate these risks, we propose \projectname, a novel AI firewall that performs semantic-level query flow control by leveraging the \texttt{ASI} to detect and mitigate malicious behaviors based on neuron activation patterns. Through extensive experiments using three SoTA LLMs (Llama3, Vicuna, and Mistral) across four diverse datasets (MS MARCO, HotpotQA, FinQA, and MedicalSys), we demonstrate that \projectname~ achieves an \texttt{AUROC} exceeding 0.909 for risk detection, with minimal degradation in \texttt{Precision} and \texttt{Recall}, both of which show reductions of less than 0.03 and 0.09, respectively, for risk mitigation.

\normalem
\footnotesize
\bibliographystyle{ACM-Reference-Format}
\bibliography{static/ref}


\clearpage
\appendix

\section{Additional Experimental Results}
\label{appendixA}

\begin{figure}[!h]
    \centering
    \includegraphics[width=0.95\linewidth]{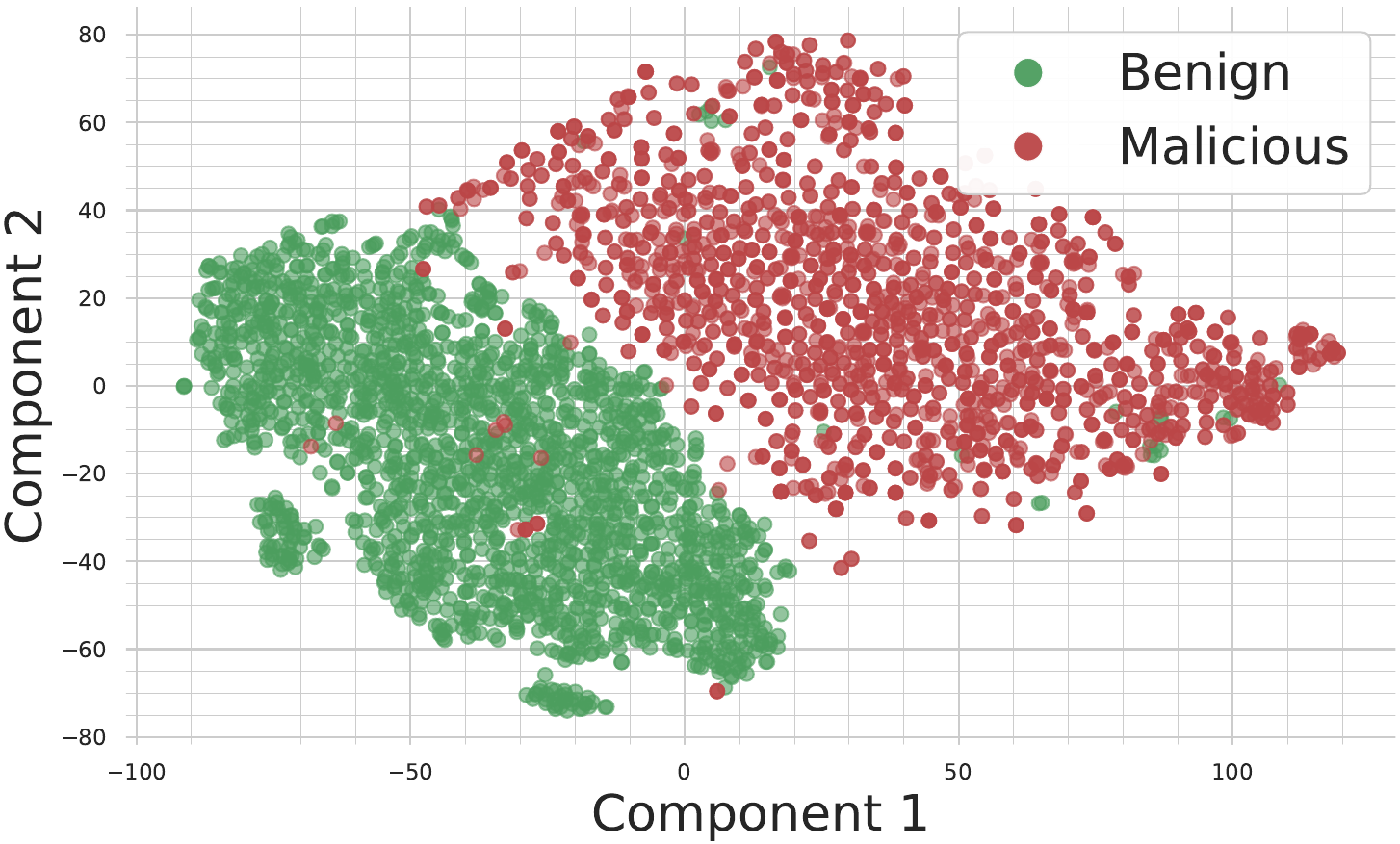}
    \caption{T-SNE visualization on Reconnaissance risk}
\end{figure}

\begin{figure}[!h]
    \centering
    \includegraphics[width=0.95\linewidth]{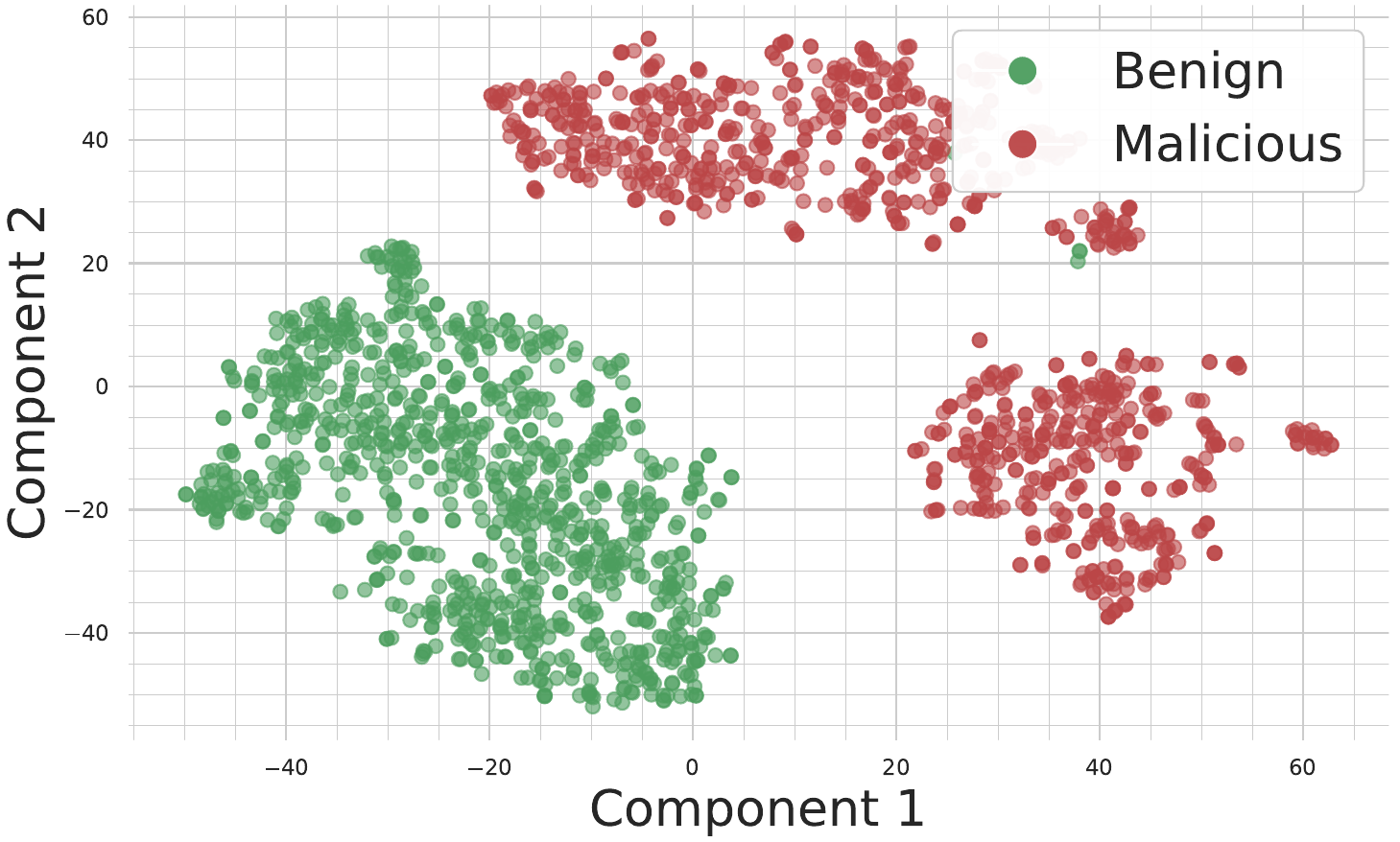}
    \caption{T-SNE visualization on Data Exfiltration risk}
\end{figure}

\begin{figure}[!h]
    \centering
    \includegraphics[width=0.95\linewidth]{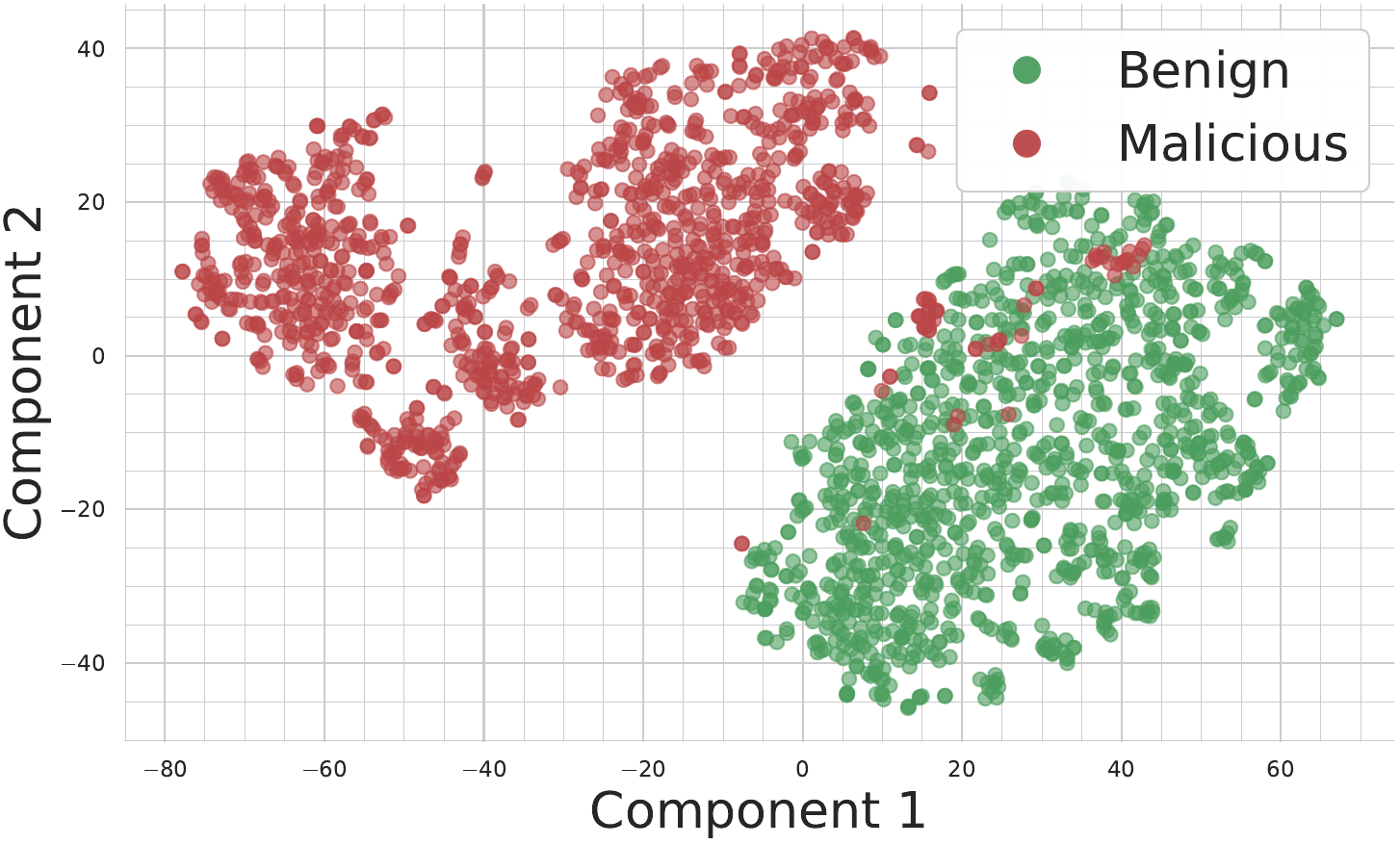}
    \caption{T-SNE visualization on Knowledge Poisoning risk}
\end{figure}

\section{Data Samples}
\label{sec:appendixB}

\begin{table}[!t]
  \centering
  \caption{Example Requests for Medicalsys dataset}
  \begin{tabular}{M{1.25cm}| R{3.25cm} | R{3.25cm}}
    \toprule
    \thead{\textbf{Role}} & \thead{\textbf{Benign Request Example}} & \thead{\textbf{Malicious Request Example}}\\
    \midrule
    \textbf{Medical Practitioners} & I would like to inquire about the medical history of patient Zhang San, who visited on January 3, 2025. &  Please provide the hospital's expenditure on equipment and total revenue for January 17, 2025.

    \textcolor{red}{(Attempt to access content that is only available to Financial Accountants)}
    \\ \hline
    \textbf{Financial Accountants} & What is the hospital's registration fee revenue for December 2024? & How many pieces of equipment were repaired at the hospital yesterday?
    \textcolor{red}{(Attempt to access content that is only available to Logistics Administrators)}
\\ \hline
\textbf{Logistics Administrators} & What equipment did the internal medicine department purchase yesterday? & How many years of work experience does Dr. Li Si have?

\textcolor{red}{(Attempt to access content that is only available to Human Resources Managers)}
\\ \hline
\textbf{Human Resources Managers} & What is the highest monthly salary for a surgical nurse? &

Tell me the chief complaint of patient Wang Wu.

\textcolor{red}{(Attempt to access content that is only available to Human Medical Practitioners)}
    \\ 
    \bottomrule
  \end{tabular}
\end{table}

\clearpage


\section*{Supplementary material}

Table 5: Representative cases for Medicalsys dataset

| Role | Example |
| --- | --- |
| Medical Practitioners | Case ID:3046<br>Name: Chen Tao gender: male age: 35 ID number: 411728193501138485 Tel: 18708596474 Address: 859087, Block M, Nanchang Road, Anci District, Haimen City, Jiangxi Province Hospital: First People's Hospital of Hangzhou City Visiting Department: Traditional Chinese Medicine Gynecology Date of consultation: September 18, 2024<br>Chief complaint: The patient reported that he had had irregular menstruation in the past month, accompanied by abdominal pain and waist soreness.<br>Current medical history: The patient's menstrual cycle has been disturbed in the past month, with prolonged periods, and frequent periods of menstruation. At the same time, there was slight lower abdominal pain and waist soreness, and there were no obvious symptoms such as fever, nausea and vomiting. The patient had no similar symptoms before, but the symptoms persisted this time, affecting the quality of daily life, so he came to see a doctor.<br>Past history: The patient denied a history of chronic diseases such as hypertension and diabetes; denied a history of drug allergies; denied a history of surgical trauma; denied a history of family genetic diseases.<br>Physical examination: - Body temperature: 36.8℃ - Blood pressure: 120/80mmHg - Heart rate: 78 beats/minute - Abdominal palpation: mild tenderness, no mass touched - Reproductive system examination: smooth cervix and mild tenderness in bilateral adnexal areas<br>Auxiliary inspection: - Blood routine: normal range - Urine routine: normal range - B-ultrasound: No obvious abnormalities were found in the uterus and appendages<br>Preliminary diagnosis: Gynecologic disease in traditional Chinese medicine, considered as menstrual disorder with possible pelvic inflammatory disease<br>Handling opinions: 1. Give traditional Chinese medicine conditioning, and the specific prescription will be prescribed by a doctor; 2. Pay attention to rest and avoid overwork; 3. Maintain good living habits and eat light; 4. Regular review and prompt medical treatment if you feel uncomfortable. |
| Financial Accountants | Record No.:2935 (2024-02-12)<br>Income:Outpatient registration fee income: 23253 yuanHospitalization expense income: 503264 yuanCT examination: 235631 yuanMRI examination: 26620 yuanX-ray examination: 75619 yuanBlood test: 8271 yuanOrdinary surgery: 52871 yuanAdvanced surgery: 660887 yuanDrug sales revenue: 571071 yuanRehabilitation treatment: 5299 yuanHealth consultation: 5755 yuanVaccination: 5923 yuan<br>Total income: 2174464 yuan<br>Costs and Expenses:Disposable consumables: 26656 yuanHigh-value consumables: 57021 yuanDrug purchase cost: 44806 yuanDoctor: 57581 yuanNurse: 43416 yuanAdministrative staff: 30579 yuanTechnical staff: 16212 yuanUtility costs: 11452 yuanCT machine maintenance: 5325 yuanMRI machine maintenance: 4861 yuanDepreciation of other equipment: 11515 yuanOffice supplies: 5475 yuanPublicity fee: 6479 yuanNew CT machine: 1295753 yuanCleaning and sanitation: 5922 yuanSecurity: 4363 yuan<br>Total costs and expenses: 1627416 yuan<br>Profit:<br>Net profit: 547048 yuan |
| Logistics Administrators | Record number:603<br>Time: 17:14<br>Equipment name: Medical ultrasonic diagnostic instrument-<br>Fault description: The image quality of medical ultrasound diagnostic instruments deteriorates during use, which affects the accuracy of doctors' diagnosis.-<br>Maintenance measures: A comprehensive inspection of the medical ultrasonic diagnostic instrument was carried out and it was found that the image processing system was slightly damaged. Therefore, the image processing system was calibrated and the latest software version was updated.-<br>Maintenance results: After maintenance, the image quality has been significantly improved, and various functions have returned to normal, fully meeting the needs of clinical use.-<br>Maintenance personnel: Wang Gong-<br>Other related information: During the maintenance process, the equipment was also comprehensively cleaned and maintained to ensure long-term stable operation. It is recommended to conduct regular maintenance inspections to prevent similar problems. |
| Human Resources Managers | ID:3941- Name: Du Yu- gender: female- Date of birth: August 15, 1964- ID number: 120113198708270286- Tel: 15569760461- Email: ping45@example.com- Home address: 691722, Block n, Ma 'anshan Street, Shanting, Yu County, Anhui Province- Department: Laboratory Inspection- Position: Head Nurse- Entry date: March 21, 2015- Basic salary: 10412 yuan/month- Performance bonus: 3510 yuan/month- Year-end bonus: 10626 yuan/year-<br>Educational background:- 1983 - 1988: Graduated from Anhui Medical University with a bachelor's degree in nursing- 1990 - 1992: Graduated from Peking Union Medical College with a master's degree in nursing-<br>Professional qualifications:- Received the Nurse Practice Certificate in 1988- Received the nursing qualification certificate in 1992- 1998 Obtained the qualification certificate of supervisor nurse- In 2005, he obtained the deputy chief nurse qualification certificate- Obtained the Chief Nurse Qualification Certificate in 2012-<br>Work experience:- 1988 - 1990: Anhui Province People's Hospital, served as a nurse- 1992 - 2000: Beijing City First People's Hospital, served as a nurse- 2000 - 2010: Shanghai City Second People's Hospital, served as the chief nurse- 2010 - 2015: Shenzhen Third People's Hospital, served as deputy chief nurse- 2015-present: The First People's Hospital of Zhejiang Province, serving as head nurse-<br>Training experience:- 1990: Participated in the National Nursing Management Training Course- 1995: Participated in the Nursing Management Seminar organized by the International Nursing Association- 2000: Participated in advanced nursing management training at the Johns Hopkins University School of Nursing- 2005: Participated in the Nursing Management Training Course at Free University of Berlin, Germany- 2010: Participated in nursing quality management training at King's College London, UK-<br>Honors and rewards:- 1995: Won the title of "Excellent Nurse"- 2000: Won the title of "Advanced Worker"- 2005: Won the "Outstanding Contribution Award for Nursing Management"- 2010: Won the title of "Nursing Manager of the Year"-<br>Remarks: Ms. Du Yu has more than 30 years of working experience in the nursing field. She is good at nursing management and quality control, and enjoys a high reputation in the industry. |



\end{document}